\newif\iflong
\newif\ifshort

\longtrue

\iflong 
\else
\shorttrue
\fi

\documentclass[11pt,letterpaper]{article}
\usepackage[lmargin=1.0in,rmargin=1.0in,bottom=1.0in,top=1.0in,twoside=False]{geometry}
\usepackage{amsmath,amssymb,graphicx,amsthm}
\usepackage{dsfont}
\usepackage{color,soul}
\usepackage{xspace}
\usepackage{enumerate}
\usepackage{mathtools}
\usepackage{bbm}

 \usepackage[numbers]{natbib}

\usepackage[pagebackref]{hyperref}
\hypersetup{%
  hypertexnames=true,
colorlinks=true,citecolor=green!60!black,linkcolor=red!60!black%
}

\usepackage{multicol}
\usepackage{booktabs}
\usepackage{paralist}

\usepackage[tt=false]{libertine}
\usepackage{inconsolata}

\usepackage{subcaption}
\usepackage[noend,ruled,linesnumbered]{algorithm2e}
\usepackage{xifthen}

\usepackage[linecolor=green!70!black, backgroundcolor=green!10, bordercolor=black, textsize=scriptsize]{todonotes}

\usepackage{mathrsfs}
\newcommand{\myemph}[1]{{\color{green!40!black}\emph{#1}}}

\newcommand{\prob}[6]{%
    \begin{description} %
      \setlength\topsep{-.15ex} \setlength\itemsep{-.2ex}
      \item[#1]
      \item[\emph{#2}]#3
      \item[\emph{#4}]#5
    \end{description}%
 }

\newcommand{\probdef}[3]{\prob{#1}{Input:}{#2}{Question:}{#3}{as}}

\usepackage{tikz}
\usetikzlibrary{decorations,arrows,petri,topaths,backgrounds,shapes,positioning,fit,calc,decorations.pathreplacing,patterns,intersections,decorations.pathmorphing,matrix}

\tikzstyle{thickline} = [line width=1.8pt]

\tikzstyle{boy} = [draw=gray!70!black, rectangle, fill=gray!70!black, inner sep=2.2pt]
\tikzstyle{girl} = [draw=gray!70!black, circle, fill=gray!70!black, inner sep=2pt]

\makeatletter
\newcommand{\gettikzxy}[3]{%
  \tikz@scan@one@point\pgfutil@firstofone#1\relax
  \edef#2{\the\pgf@x}%
  \edef#3{\the\pgf@y}%
}
\makeatother

\newcommand{\dop}{divorce}
\newcommand{\dopm}{\mathsf{div}}
\newcommand{\divorcesm}{\textsc{Reaching Stable Marriage via Divorces}}
\newcommand{\divorcesms}{\textsc{DivorceSM}}

\def \xdist {4ex}
\def \ydist {3ex}

\usepackage{cleveref}

\newtheorem{corollary}{Corollary}
\newtheorem{lemma}{Lemma}

\newtheorem{theorem}{Theorem}
\newtheorem{claim}{Claim}

\theoremstyle{definition}

\newtheorem{example}{Example}

\crefname{table}{Table}{Tables}
\crefname{figure}{Figure}{Figures}
\crefname{theorem}{Theorem}{Theorems}
\crefname{definition}{Definition}{Definitions}
\crefname{corollary}{Corollary}{Corollaries}
\crefname{observation}{Observation}{Observations}
\crefname{lemma}{Lemma}{Lemmas}
\crefname{example}{Example}{Examples}
\crefname{reduction}{Reduction}{Reductions}
\crefname{construction}{Construction}{Constructions}
\crefname{subsection}{Subsection}{Subsections}
\crefname{section}{Section}{Sections}
\crefname{proposition}{Proposition}{Propositions}
\crefname{algorithm}{Algorithm}{Algorithms}
\crefname{claim}{Claim}{Claims}

\newcommand{\cqed}{$\diamond$}

\newcommand{\vvv}{\color{blue!70!black} v}
\newcommand{\ww}{\color{red!70!black} w}
\newcommand{\yy}{\color{red!70!black} y}
\newcommand{\xx}{\color{blue!70!black} x}
\newcommand{\nvv}{\color{blue!70!black} \overline{v}}
\newcommand{\nww}{\color{red!70!black} \overline{w}}
\newcommand{\nyy}{\color{red!70!black} \overline{y}}
\newcommand{\nxx}{\color{blue!70!black} \overline{x}}
\newcommand{\ff}{\color{green!70!black} f}
\newcommand{\pp}{\color{green!70!black} p}
\newcommand{\ee}{\color{orange!70!black} e}
\newcommand{\qq}{\color{orange!70!black} q}
\newcommand{\mysucc}{\;}
\newcommand{\kk}{\ensuremath{k}}
\newcommand{\true}{\mathsf{true}}
\newcommand{\false}{\mathsf{false}}
\newcommand{\lit}{\ensuremath{\mathsf{lit}}}

\newcommand{\wonehardresult}{%
   \divorcesms{} is W[1]-hard with respect to the total number~$\kappa$ of allowed \dop{s}.%
 }

\newcommand{\bps}{\ensuremath{\mathsf{bps}}}

\newcommand{\pspace}{\text{\normalfont PSPACE}}
\newcommand{\npspace}{\text{\normalfont NPSPACE}}
\newcommand{\np}{\text{\normalfont NP}}

\newcommand{\callcheckBP}{\textsf{checkBP}}
\newcommand{\callMain}{\textsf{main}}

\title{Reaching Stable Marriage via Divorces is Hard}
\author{Jiehua Chen \\
  {TU Wien, Vienna, Austria}}
\date{}

\begin{document}

\maketitle

\begin{abstract}
  We study the \divorcesm{}~(\divorcesms) problem of deciding,
  given a \textsc{Stable Marriage} instance and an initial matching~$M$,
  whether there exists a stable matching which is reachable from~$M$ by divorce operations as introduced by Knuth~\cite{Knuth1976}. 
  Towards answering an open question of Manlove~\cite{Manlove2013} and Cechl{\'a}rov{\'a} et al.~\cite{CCM2019-openq},
  we show that for incomplete preferences without ties, \divorcesms{} is NP-hard.
  Our hardness reduction also implies that the problem remains parameterized intractable for the number~$\kappa$ of allowed divorce operations.
  It remains NP-hard even if the maximum length~$d$ of the preferences is a constant.
  For the combined parameter~$(\kappa, d)$, the problem is fixed-parameter tractable.
\end{abstract}

\section{Introduction}

In the \textsc{Stable Marriage} problem, we are given %
two disjoint sets of agents, $U$ and $W$, of equal size,
which we refer to as to the set of men and the set of women, respectively.
Each agent  %
has a strict preference list over all agents of the opposite sex. %
The goal is to compute a \myemph{matching~$M$} which is \myemph{stable}.
Herein, a \myemph{matching} is a disjoint set of man-woman pairs.  %
Matching~$M$ is \myemph{stable} if no unmatched pair is blocking~$M$.
A pair~$\{u,w\}$ of agents is \myemph{blocking} matching~$M$ if both~$u$ and $w$ prefer being paired together over the outcome given by matching~$M$. 

Gale and Shapley~\cite{GaleShapley1962} showed that an instance of the \textsc{Stable Marriage}~(\textsc{SM}) problem with $n$ men and $n$ women always admits a stable matching and provided an $O(n^2)$-time algorithm to find one.
Their algorithm is, however, static and any changes in agents' preferences will make the algorithm start from scratch again.
In this case, a natural approach, without disturbing the existing matching too much, is to iteratively find a blocking pair~$\{u,w\}$ and match~$u$ and $w$ and their divorcees together, respectively.
We call the procedure in each such iteration a \myemph{divorce operation} (in short \myemph{divorce}).
Knuth~\cite{Knuth1976} provided a small example to demonstrate that not every divorce operation can result in success, i.e., not every possible sequence of divorce operations can reach a stable matching.
He proposed two interesting questions:
\begin{quote}
  \emph{Is there a successful sequence of divorce operations from any matching, ultimately reaching a stable matching? If so, is there one that is relatively short?}
\end{quote}
In their influential book, Gusfield and Irving~\cite{GusfieldIrving1989} also asked whether the relation of the divorce operations may have some specific structure that helps in identifying a sequence of divorces leading to a stable matching. 

Independently, Tamura~\cite{Tamura1993} and Tan and Su~\cite{TanSu1995} answered the first question by Knuth~\cite{Knuth1976} negatively.
More precisely, they showed that there are matchings for which no sequence of divorce operations leads to any stable matching.
Despite these negative findings, %
from a computer science point of view, and in an attempt to answer the second question of Knuth,
it would still be interesting to know whether, and if so, how, a given matching can reach a stable one via divorce operations.
This question is considered as one of the fundamental and most intriguing problems in algorithmics for matching under preferences, and its computational complexity remains open~\cite{Manlove2013,HoeferWagner2017-local-sm,CCM2019-openq}.
Towards answering this open question,
in this paper, we investigate the algorithmic complexity of the underlying computational problem, called \divorcesm{}~(\divorcesms); see \cref{sec:preli} for the formal definition. %

\looseness=-1
\paragraph{Our contributions.}
It is straight-forward to see that \divorcesms{} is in the complexity class \npspace{}, and hence~\pspace{},
since one can, for each~$i\in \{0,1,\ldots,\}$, 
iteratively guess a man-woman pair~$\rho_i$ and check, using polynomial space, whether $\rho_i$ is blocking $M_i$, compute a next matching~$M_{i+1}$ using the divorce operation on~$\rho_i$,
and check whether $M_{i+1}$ is stable in polynomial time~(see \cref{cor:PSPACE}).
It is, however, open whether \divorcesms{} belongs to \np{} since it is unclear whether the length of any successful sequence of the divorce operations has polynomial size.
Nevertheless, we can show NP-hardness, establishing the first computational complexity lower bound for \divorcesms{}:

\begin{theorem}\label{thm:divor-incomplete-noties-nphard}
  \divorcesms{} for incomplete and strict preferences is \np-hard. %
\end{theorem}

This hardness result, albeit for incomplete preferences, is surprising since, besides a few exceptions~\cite{Kato1993,Feder1992a,CheRoSo2020}, most problems in the context of \textsc{Stable Marriage} with strict preferences are polynomial-time solvable.

\looseness=-1
To understand the impact of the length of a sequence of the \dop{s}, i.e., when a stable matching is reachable after few \dop{s},
or when the length of the preference lists is small, 
we further study the parameterized complexity of \divorcesms{}, that is, two restricted variants of the problem.
Clearly, \divorcesms{} can be solved in polynomial time if the number of \dop{s} in a successful sequence (if any) is a constant: For each possible sequence of blocking pairs, check whether performing the corresponding divorce operations will reach a stable matching.
From the parameterized complexity point of view, this means that \divorcesms{} is in XP with respect to the parameter ``maximum number~$\kappa$ of \dop{s}'' in a successful sequence~(see \cref{cor:dop-xp}).
Hence, it is tempting to ask whether we can improve on the running time by providing a fixed-parameter algorithm.
Our next result answers this negatively.

\begin{theorem}\label{thm:dop-w1hard}
 \wonehardresult
\end{theorem}

As for the length of the preference lists, by reducing from a restricted variant of the NP-complete \textsc{3SAT} problem, we show the following.

\begin{theorem}\label{thm:NP-hard-constant-length}
  \divorcesms{} remains \np-hard even if each preference list has constant length.
\end{theorem}

Combining both the number~$\kappa$ of allowed \dop{s} and the maximum length~$d$ of the preference lists, we obtain fixed-parameter tractability.
\begin{theorem}\label{thm:FPT}
  \divorcesms{} can be solved in $O((4d)^{\kappa}\cdot \kappa! \cdot n^2)$~time.
\end{theorem}

\paragraph{Related work.}
As already mentioned, not every matching can be transformed into a stable matching via \dop{s}. Tamura~\cite{Tamura1993} provided a construction that, for each number~$n\ge 4$, produces an instance with $n$~agents on each side,
for which there is a matching that does not lead to a stable matching by performing any sequence of \dop{s}.
The instance with $n=4$ is depicted in \cref{ex1} (also see \cite[Fig.~4]{Tamura1993}).
He also provided an algorithm that transforms an arbitrary matching into a stable one by using operations that are not only \dop{s}.
His algorithm does not necessarily run in polynomial time.
Independently, Tan and Su~\cite{TanSu1995} provided a different instance with four agents on each side where there is a matching which also cannot be transformed into a stable matching using only divorces.
They also showed that for any instance with at most three agents on each side and with complete preferences but without ties, an arbitrary matching can always be transformed into a stable one by using only \dop{s}.
Similarly, they also provided an algorithm that transforms an arbitrary matching into a stable one while not exclusively using \dop{s}.
For instance, they do not require the divorcees to be matched together as we do in this paper and as originally defined by Knuth.
This changes the situation dramatically, since if the divorcees may be left unmatched, then every matching can be transformed to a stable matching by forcing the agents in blocking pairs to be matched together. 
See the following work~\cite{RothVVate1990,AbeRot1995,AckGolMirRoVo2011,Cheng2016} and the textbook of Manlove~\cite[Section 2.6]{Manlove2013} for more details on this setting.

\paragraph{Organization of the paper.}
In \cref{sec:preli}, we introduce relevant concepts and notations.
In \cref{sec:NP-hard}, we prove \cref{thm:divor-incomplete-noties-nphard}. %
In \cref{sec:param}, we consider the parameterized complexity with regard to \iflong two prominent parameters,
\fi the number of divorces and the maximum length of the preferences of each agents,
and prove \cref{thm:dop-w1hard,thm:NP-hard-constant-length,thm:FPT}. 
In \cref{sec:conclude}, we conclude with several open questions.
Due to space constraints, some proofs are deferred to the appendix.

\section{Preliminaries}\label{sec:preli}
Given a non-negative integer~$z$, we use \myemph{$[z]$} to denote the set~$\{1,2,\ldots,z\}$.

\paragraph{Stable Marriage~(SM).} An instance~$I=(U,W,(\succ_x)_{x\in U\cup W})$ of \textsc{SM} consists of two $n$-element disjoint sets of agents, $U$ and $W$ such that for each agent~$u\in U$, the notation~\myemph{$\succ_u$} denotes a linear order on a subset~$W'$ of~$W$ that represents the ranking of agent~$u$ over all agents from $W'$.
The agents in~$W'$ are also called \myemph{acceptable} to~$u$.
The agents \emph{not ranked} in $\succ_u$ are those in $W \setminus W'$, that is, those that $u$ does not agree to be matched with; we also call them \myemph{unacceptable}.
If $w \succ_u w'$, then we say that $w$ is \myemph{preferred} to~$w'$ by~$u$.
Analogously, for each agent~$w \in W$, $\succ_w$ represents a linear order on a subset of~$U$ that represents the ranking of~$w$ and we likewise use the notions of preference list, preferred, and (un)acceptable.
We assume that the acceptability relation among the agents is \emph{symmetric}, i.e.,
for each two agents~$u$ and $w$ it holds that $u$ is acceptable to~$w$ if and only if $w$ is acceptable to~$u$.
We say that instance~$I$ has \myemph{complete} preferences if each agent finds all agents of the opposite set acceptable; otherwise it has \myemph{incomplete} preferences.

A matching of~$I$ is a set of  pairwise disjoint pairs, each pair containing one agent from~$U$ and one agent from $W$, {i.e.}, 
$M\subseteq \{\{u,w\}\mid u \in U \wedge w \in W\}$ and for each two distinct pairs~$p, p'\in M$ it holds that $p \cap p' = \emptyset$.
Given a pair~$\{u,w\}$ with $u\in U$ and $w\in W$,  
if it holds that $\{u,w\}\in M$, then we use \myemph{$M(u)$} to refer to $w$ and \myemph{$M(w)$} to refer to $u$,
and we say that $u$ and $w$ are their respective \myemph{partners} under~$M$;
otherwise we say that $\{u,w\}$ is an \myemph{unmatched pair} under~$M$.
If an agent~$x$ is \emph{not} assigned any partner by $M$, then we say that $x$ is \myemph{unmatched by~$M$}. %

We say that a pair~$\{u,w\}$ is \myemph{blocking} (or \myemph{a blocking pair of}) \myemph{$M$} if %
the following holds:
\iflong \begin{enumerate}[(i)]
  \item
  \else
  (i)
  \fi $u$ and $w$ find each other acceptable but are not matched to each other,
  \iflong \item \else (ii) \fi $u$ is either unmatched by $M$ or $u$ prefers~$w$ to~$M(u)$, and
  \iflong \item \else (iii) \fi $w$ is either unmatched by $M$ or $w$ prefers~$u$ to~$M(w)$.
\iflong \end{enumerate}\fi
Finally, we say that a matching~$M$ is \myemph{stable} if it does not admit a blocking pair. %

\paragraph{Reachable matchings and divorce operations.}
For notational convenience, given a matching~$M$ and a pair~$\rho=\{u,w\}$ of agents with~$\{u,w\}\notin M$,
we use \myemph{$\dopm(M, \rho)$} to denote the set resulting from replacing the pairs~$\{u,M(u)\}$ and $\{M(w), w\}$ in matching~$M$ with $\{u,w\}$ and $\{M(w), M(u)\}$,
while keeping the other pairs unchanged.
Formally,
\begin{align*}
  \dopm(M, \rho) \coloneqq \Big(M \setminus \{\{u, M(u)\}, \{M(w), w\}\}\Big) \cup  \{\{u, w\},
  \{M(w), M(u)\}\}.
\end{align*}
If $\rho$ is blocking $M$ and the set~$\dopm(M, \rho)$ is a matching of~$I$, then the above operation is called a \myemph{\dop{} by $\rho$ for $M$}. %

Given two matchings~$M_0$ and $M$ of an SM instance~$I=(U,W,(\succ_x)_{x\in U\cup W})$,
we say that $M$ is \myemph{reachable from~$M_0$} if there exists  a \myemph{sequence~$L=(\rho_0,\rho_1,$ $\ldots, \rho_{\ell-1})$} of acceptable pairs of agents, where $\rho_i=\{u_i,w_i\}$ for all~$i\in \{0,1,\ldots,\ell-1\}$, satisfying the following:
\begin{itemize}[--]
  \item For each~$i\in\{1,2,\ldots,\ell\}$, let $M_{i}$ be a matching of $I$, recursively defined as
   \begin{align}
     M_{i+1} \coloneqq \dopm(M_{i}, \rho_{i}).\label{eq:divorce}
   \end{align}
  \item Each pair~$\rho_i$ is blocking matching~$M_{i}$, and
  \item $M=M_\ell$.
\end{itemize}
We also call $L$ a \myemph{witness} that~$M_\ell$ is reachable from~$M_0$.

\begin{example}\label{ex:incomplete-pref}
  For an illustration, let us consider the following instance with three agents on each side: $U=\{a,b,c\}$ and $W=\{1,2,3\}$.
  Their preference lists are depicted below; throughout, we omit the subscript~$x$ in the the preference list~$\succ_x$ for the sake of brevity.
  \begin{align*}
    \begin{array}{l@{\,}lcl@{\,}l}
      a\colon& 2 \succ 3 \succ \boxed{1}, &~~~ &1\colon& \boxed{a} \succ b \succ {c},\\
      b\colon& 1 \succ 3 \succ \boxed{2}, &~~~ &2\colon& \boxed{b} \succ a,\\
      c\colon& \boxed{3} \succ 1, &~~~ &3\colon& b \succ a \succ \boxed{c}.
    \end{array}
  \end{align*}
  The initial matching~$M_0$ with~$M_0=\{\{a,1\}, \{b,2\}, \{c,3\}\}$ is marked with boxes in the above preference lists and 
  has two blocking pairs~$\rho_1=\{a,3\}$ and $\rho_2=\{b,3\}$.
  We can perform a \dop{} by~$\rho_1$ but not by~$\rho_2$ since the divorcees of the agents in~$\rho_2$, namely $M_0(b)=2$ and $M_0(3)=c$, do not find each other acceptable.
  This means that $\dopm(M_1,\rho_1)$ is a valid matching while $\dopm(M_1, \rho_2)$ is not.
  After we perform a \dop{} by~$\rho_1$, we obtain matching~$M_2=\{\{a,3\}, \{b,2\}, \{c,1\}\}$, which has two blocking pairs, namely~$\rho_2$ and $\rho_3$ with $\rho_3=\{b,1\}$.
  Note that $\dopm(M_2,\rho_2)$ is a valid matching while $\dopm(M_2,\rho_3)$ is not.
  Now, we can perform a \dop{} by~$\rho_2$ to obtain matching~$M_3$ with~$M_3=\{\{a,2\}, \{b,3\}, \{c,1\}\}$.
  $M_3$ still has a blocking pair~$\rho_4=\{b,1\}$ by which performing a \dop{} will result in the unique stable matching~$M_4$ with $M_4=\{\{a,2\}, \{b,1\}, \{c,3\}\}$.
\end{example}

The following example by Tamura~\cite{Tamura1993} illustrates the situation where a matching can never reach stability.
\begin{example}\label{ex1}
Consider the following instance with four agents on each side\iflong : $U=\{u_1,u_2,u_3,u_4\}$ and $W=\{w_1,w_2,w_3,w_4\}$.\fi.
\begin{multicols}{2}
\begin{tabular}{lllllll}
    $u_1\colon$ & $\boxed{w_1} \succ w_3 \succ w_2 \succ w_4$\\
    $u_2\colon$ & $\boxed{w_2} \succ w_4 \succ w_3 \succ w_1$\\
    $u_3\colon$ & ${w_3} \succ w_1 \succ \boxed{w_4} \succ w_2$\\
    $u_4\colon$ & ${w_4} \succ w_2 \succ w_1 \succ \boxed{w_3}$\\
\end{tabular}

\columnbreak
\begin{tabular}{lllllll}
    $w_1\colon$ & $u_2 \succ u_4 \succ \boxed{u_1} \succ u_3$\\
    $w_2\colon$ & $u_3 \succ u_1 \succ \boxed{u_2} \succ u_4$\\
    $w_3\colon$ & $\boxed{u_4} \succ u_2 \succ {u_3} \succ u_1$\\
    $w_4\colon$ & $u_1 \succ \boxed{u_3} \succ {u_4} \succ u_2$\\
\end{tabular}
\end{multicols}

If we start with matching~$N_0=\{\{u_1,w_1\}, \{u_2,w_2\}, \{u_3,w_4\}, \{u_4,w_3\}\}$ (see the agents in boxes),
then no matter what \dop{s} we perform, we will never reach a stable matching.
For more details, please refer to the work of Tamura~\cite{Tamura1993}.
\end{example}

\paragraph{Central problem.}
\noindent The problem that we study in this paper is formally defined as follows.
\probdef{\divorcesm~(\divorcesms)}
{An instance~$I=(U, W, (\succ_{x})_{x\in U\cup W})$ of the SM problem (possibly with incomplete preferences) and an initial matching~$M_0$ of~$I$.}
{Does $I$ admit a stable matching which is reachable from $M_0$?}

Checking whether there exists a witness with $\ell$~pairs for~$M_0$ can be done in $O(\ell \cdot n^2)$ time, implying that \divorcesms{} is in \pspace.

\begin{lemma}\label{lem:check-witness}
  Let~$I=(U,W,(\succ_{x})_{x\in U\cup W}, M_0)$ be a \divorcesms{} instance
  and let $L$ be a sequence of~$\ell$ pairs of agents from~$I$.
  Checking whether~$L$ is a witness for~$M_0$ can be done in $O(\ell\cdot n^2)$~time and in $O(n^2)$~space.
\end{lemma}

\begin{proof}
  Let $I$ and $L$ be as defined with $L=(\rho_0,\rho_1,\ldots, \rho_{\ell})$.
  To show the desired space complexity, we provide an algorithm which works in $\ell$ iterations.
  In each iteration, it only processes one pair from~$L$.
  More precisely, for each $i\in \{0,1,\ldots,\ell-1\}$, let $\rho_{i}=\{u_i,w_i\}$ and we do the following:
  \begin{itemize}[--]
    \item Check whether $\rho_{i}$ is a blocking pair of~$M_{i}$, and $M_i(u_i)$ and $M_i(w_i)$ find each other acceptable in constant time.
    \begin{enumerate}[(a)]
      \item Reject $L$ if the check returns no;
      \item Otherwise, construct~$M_{i+1}\coloneqq \dopm(M_{i}, \rho_i)$  in $O(n^2)$ time, and proceed with the next iteration.
    \end{enumerate}
    \item Check whether $M_\ell$ is stable in $O(n^2)$~time and accept $L$ if and only if $M_{\ell}$ is stable.
  \end{itemize}
  The running time of the above procedure is also straight-forward to check.
  Since each matching needs $O(n)$ space,
  and the matchings of previous iterations are not relevant to the current iteration, we can reuse this $O(n)$ space when we compute the next matching.
  The input instance~$I$ needs $O(n^2)$ space.
  Hence, checking whether a given sequence is a witness needs $O(n^2)$ space.
\end{proof}

\cref{lem:check-witness} implies the following upper bound for our problem.

\begin{corollary}\label{cor:PSPACE}
  \divorcesms{} is contained in \pspace.
\end{corollary}

\iflong
\begin{proof}
  To show \pspace-containment, we show that \divorcesms{} is contained in  \npspace{} since \pspace{}$=${}\npspace~\cite{Pap94}.
  To show \npspace-containment, we provide a non-deterministic algorithm that always terminates,
  uses polynomial-space, and decides \divorcesms{}. 

  Let $I=(U,W,(\succ_x)_{x\in U\cup W}, M_0)$ be an instance of \divorcesms{} with $|U|=|W|=n$.
  Assume that $I$ is a yes-instance, and let $L=(\rho_0,\rho_1,\ldots,\rho_{\ell})$ be a shortest witness for a stable matching~$M$ to be reachable from~$M_0$.
  Let $L'=(M_0,M_1,\ldots,M_{\ell}=M)$ be the sequence of the corresponding matchings, defined according to~\eqref{eq:divorce}.
  Using \cref{lem:check-witness}, we can iteratively guess the blocking pair~$\rho_i$ in~$L$ and check,
  using $O(n^2)$ space, whether $\rho_i$ is blocking~$M_i$ and whether $\dopm(M_i, \rho_i)$ is a matching of $I$, and then define~$M_{i+1}=\dopm(M_i,\rho_i)$.
  Since $L$ is a shortest witness, it follows that each~$(M_i,\rho_i)$, $i\in \{0,\ldots,\ell\}$ is distinct. 
  Since there are at most~$n!$ matchings and $n^2$ man-woman pairs, 
  it follows that $\ell \le n!\cdot n^2$.
  This means that after at most~$n!\cdot n^2$ guesses, we will be able to confirm that $I$ is a yes-instance by finding a shortest witness~$L$, but without storing the whole sequence.

  By the above reasoning, if $I$ is a no-instance, then after at most $n!\cdot n^2$ guesses, we can confirm that no sequence of blocking pairs can reach stability and return no.
  Hence, our approach always terminates (after at most $n!\cdot n^2$ steps), uses polynomial space, and gives a correct answer to each input.
\end{proof}
\fi

\section{NP-hardness for \divorcesms{}}\label{sec:NP-hard}
In this section, we %
provide the first evidence that reaching stability via divorces is intractable even if the preferences are strict but may be incomplete. %
We reduce from the NP-complete problem~\textsc{Clique}~\cite{GJ79}.
This problem is to decide, given a graph~$G=(V,E)$ with $V=\{v_1,\ldots, v_n\}$ and $E=\{e_1,\ldots, e_{m}\}$ being the vertex set and the edge set, respectively,
and a non-negative integer~$h$, whether there exists a size-$h$ clique, i.e., a vertex subset~$V'\subseteq V$ of size~$h$ such that the subgraph induced by~$V'$ is complete.

The main force behind our reduction are two seemingly similar gadgets, the vertex-gadget and the edge-gadget,
together with a vertex-selector gadget with $O(h)$ agents and an edge-selector gadget with $O(n^2)$~agents. %
Without the selector gadgets, the initial matching is a stable matching, optimal for one side of the agents.
However, due to these selector gadgets, any successful sequence of divorces must identify $\binom{h}{2}$ edge-agents and $h$ vertex-agents which correspond to a clique of size~$h$.
More specifically, the main idea behind these gadgets is as follows:
\begin{itemize}[--]
  \item There are four groups of edge-agents, called $E$, $F$, $P$, $Q$, each of size~$m$,
  which one-to-one correspond to the edges in~$E$.
  There are four groups of edge-selector-agents, called $C$, $D$, $R$, $Z$, each of size~$\binom{h}{2}$.
  Initially, each edge-selector agent from~$D$ forms with each edge-agent from~$F$ a blocking pair.
  However, ``resolving'' the blocking pairs involving the edge-selector agents
  will ultimately make~$\binom{h}{2}$ edge-agents from~$E$ \myemph{unhappy}, meaning that
  each of such edge-agents, say~$e_j\in E$, will form with \emph{each} of its two ``incident'' vertex-agents (i.e., agents which correspond to the endpoints of the edge~$e_j$) a blocking pair.
  
  \item There four groups of vertex-agents, called~$V$, $X$, $W$, $Y$, each of size~$n$,
  which one-to-one correspond to the vertices in~$V$.
  There are four groups of vertex-selector-agents, called~$S$, $T$, $A$, $B$, each of size~$h$,
  Simultaneously, each vertex-selector agent from $S$ forms with each vertex-agent from~$V$ a blocking pair.
  Each~$s_{\kk}\in S$ can only be used once to ``resolve'' one blocking pair involving some agent from~$V$.
  Resolving the blocking pairs involving the vertex-selector agents  will ultimately make \emph{exactly~$h$} vertex-agents from~$V$ \myemph{happy}, meaning that
  each of such vertex-agents will \emph{not} form with any of its ``incident'' edge-agents a blocking pair.

  \item Due to the preferences of the edge-agents and the vertex-agents, for each unhappy edge-agent, both endpoints of the corresponding edge must correspond to the happy vertex-agents.
  By our reasoning above, for each possible sequence of divorces, at least~$\binom{h}{2}$ edge-agents will become unhappy, but only $h$ vertex-agents become happy.
  Hence, a stable matching is reached only if the corresponding vertices form a clique of size~$h$.
\end{itemize}
The fact that the preferences may be incomplete helps to avoid some unexpected divorce operations.
However, the main difficulty of the reduction lies in constructing strict preferences so that we indeed can reach a desired stable matching.
Moreover, the vertex gadget and the edge gadget are designed in such a way that if the input graph admits a clique of size~$h$, then we can reach a stable matching via a sequence of~$O(h^2)$ \dop{s}.
As a final remark, if a stable matching is reachable, then there are indeed exponentially many witnesses for the reachability, each of length~$O(h^2)$.

\begin{proof}[Proof of \cref{thm:divor-incomplete-noties-nphard}]
  As said, we reduce from the NP-complete \textsc{Clique} problem~\cite{GJ79}. %

  Let $I=(G=(V,E), h)$ be an instance of \textsc{Clique} with $V=\{v_1,\ldots, v_n\}$ and $E=\{e_1,\ldots,e_m\}$.
  Without loss of generality, assume that $1 < h < n$.
  Our \divorcesms{} instance consists of two disjoint sets of agents, $\hat{U}$ and $\hat{W}$,
  with $\hat{U} \coloneqq V\uplus X \uplus T \uplus A \uplus F\uplus P \uplus C \uplus R$
  and $\hat{W} \coloneqq W\uplus Y \uplus S \uplus B \uplus E \uplus Q \uplus D \uplus Z$, where
  
  \noindent
  \begin{tabular}{@{}c@{\;}llll}
   -- & $X \coloneqq \{{\xx_i} \mid i\in [n]\}$, & $W \coloneqq \{{\ww_i} \mid i\in [n]\}$, &  $Y\coloneqq \{{\yy_i} \mid i \in [n]\}$,\\
    -- & $F \coloneqq \{{\ff_j} \mid j \in [m]\}$, & $P\coloneqq \{{\pp_j} \mid j\in [m]\}$, & $Q\coloneqq \{{\qq_j} \mid j\in [m]\}$,\\
    -- & $S \coloneqq \{s_\kk \mid \kk \in [h]\}$,  &$T\coloneqq \{t_\kk \mid \kk \in [h]\}$, & $A\coloneqq \{a_{\kk} \mid {\kk} \in [h]\}$, &  $B\coloneqq \{b_{\kk} \mid {\kk} \in [h]\}$,\\
    -- & $C\coloneqq \{c_{\kk} \mid {\kk} \in [\binom{h}{2}]\}$, & $D\coloneqq \{d_{\kk} \mid {\kk} \in [\binom{h}{2}]\}$, & $R\coloneqq \{r_{\kk} \mid {\kk} \in [\binom{h}{2}]\}$, &$Z\coloneqq \{z_{\kk} \mid {\kk} \in [\binom{h}{2}]\}$.
  \end{tabular}

  \noindent  In words:
  \begin{itemize}[--]
    \item For each vertex~$v_i \in V$, we introduce a \myemph{vertex gadget} consisting of four vertex-agents~${\vvv_i},{\xx_i},{\ww_i},{\yy_i}$.
    \item For each edge~$e_j\in E$, we introduce an \myemph{edge gadget} consisting of four edge-agents~${\ff_j}, {\pp_j}, {\ee_j}, {\qq_j}$.
    \item For each~${\kk} \in [h]$, we also introduce four \myemph{vertex-selector-agents}~$t_{\kk},s_{\kk},a_{\kk}, b_{\kk}$ who shall ``deviate'' with the agents corresponding to the ``clique'' vertices.
    \item  Finally, for each~${\kk}\in [\binom{h}{2}]$, we introduce four further \myemph{edge-selector-agents}~$c_{\kk},d_{\kk},r_{\kk},z_{\kk}$ who shall ``deviate'' with the agents corresponding to the ``clique'' edges. 
  \end{itemize}
  Note that to enhance the connection between the vertices and their corresponding vertex-agents (resp.\ the edges and their corresponding edge-agents) we use the same symbol~$v_i$ (resp.\ $e_j$) for both the vertex (resp.\ the edge) and the corresponding vertex-agent (resp.\ the corresponding edge agent).  In total, we have introduced $2n+2h+2m+2\binom{h}{2}$ agents on each side.
  
  The preferences of the agents are depicted in \cref{fig:prefs}; we omit the ``preferring'' symbol~$\succ$ for brevity.
  \begin{figure}[t!]
    \begin{align*}
      \begin{array}{l@{\;}r@{\,}lr@{\,}l}
        \forall i \in [n]\colon   & {\vvv_i} \colon & {\ww_i} \mysucc [{\color{orange!70!black}E(v_i)}] \mysucc s_1\mysucc \cdots \mysucc s_{h} \mysucc \boxed{\yy_i},
        & {\ww_i} \colon & \boxed{\xx_i} \mysucc {\vvv_i} \mysucc t_1\mysucc \cdots \mysucc t_{h}, \\
        \forall i \in [n]\colon   &  {\xx_i} \colon & {\yy_i}\mysucc b_1\mysucc \cdots \mysucc b_{h}\mysucc \boxed{\ww_i},
        & {\yy_i}\colon & \boxed{\vvv_i} \mysucc {\xx_i} \mysucc a_1\mysucc \cdots \mysucc a_{h},\\
        \forall \kk \in [h]\colon   &  {t_{\kk}}\colon & \boxed{b_{\kk}} \mysucc s_{\kk} \mysucc {\ww_1}\mysucc \cdots \mysucc {\ww_n},
        & s_{\kk} \colon & {t_{\kk}} \mysucc {\vvv_1} \mysucc \cdots \mysucc  {\vvv_n} \mysucc \boxed{a_{\kk}}, \\
        \forall \kk \in [h]\colon   & {a_{\kk}}\colon & \boxed{s_{\kk}} \mysucc b_{\kk} \mysucc {\yy_1} \mysucc \cdots \mysucc  {\yy_n}, 
         & b_{\kk} \colon &  a_{\kk}  \mysucc {\xx_1} \mysucc \cdots \mysucc {\xx_n}  \mysucc \boxed{t_{\kk}},\\[2ex]
        \forall j \in [m]\colon   &        {\ff_j}\colon & {\qq_j} \mysucc d_1 \mysucc \cdots \mysucc d_{\binom{h}{2}} \mysucc \boxed{\ee_j},
        & {\ee_j}\colon & \boxed{\ff_j}\mysucc {\vvv_i} \mysucc {\vvv_{i'}}\mysucc {r_1} \mysucc \cdots \mysucc {r_{\binom{h}{2}}} \mysucc {\pp_j}\mysucc c_1\mysucc \cdots \mysucc c_{\binom{h}{2}},\\
       \forall j \in [m]\colon   & {\pp_j}\colon & {\ee_j} \mysucc \boxed{\qq_j},
        & {\qq_j}\colon & \boxed{\pp_j}  \mysucc c_1\mysucc \cdots \mysucc c_{\binom{h}{2}} \mysucc {\ff_j},\\
 \forall \kk \in [\binom{h}{2}]\colon   &        {c_{{\kk}}}\colon & \boxed{z_{{\kk}}} \mysucc {\ee_1}\mysucc \cdots \mysucc {\ee_m} \mysucc d_{{\kk}} \mysucc {\qq_1}  \mysucc \cdots \mysucc {\qq_m},~~~~~
        &  {d_{{\kk}}}\colon &  c_{{\kk}} \mysucc {\ff_1} \mysucc \cdots \mysucc {\ff_m} \mysucc \boxed{r_{{\kk}}},\\
   \forall \kk \in [\binom{h}{2}]\colon   &      {r_{{\kk}}} \colon& \boxed{d_{{\kk}}} \mysucc z_{{\kk}} \mysucc {\ee_1}\mysucc \cdots \mysucc{\ee_m},
        &   {z_{{\kk}}}\colon & r_{{\kk}}\mysucc \boxed{c_{{\kk}}}. \\
      \end{array}
    \end{align*}
    \caption{Preferences of the agents constructed in the proof of \cref{thm:divor-incomplete-noties-nphard}.
      Here, ${\color{orange!70!black}E(v_i)}=\{{\ee_j}\mid v_i \in e_j \in E\}$ denotes the set of edge-agents corresponding to the edges which are incident to the vertex~$v_i$ and
      $[E(v_i)]$ denotes an arbitrary but fixed order of the agents in~$E(v_i)$.
      The two vertex-agents~${\vvv_i}$ and ${\vvv_{i'}}$ in the preference list of edge-agent~$e_j$ correspond to the endpoints~$v_i$ and $v_{i'}$ of edge~$e_j$ with $i<i'$.
      In the initial matching, each agent is matched with the one marked in a box.
      For instance,
       for each $i\in [n]$,
       \iflong vertex-agent~{$\vvv_i$}
       \else {$\vvv_i$}
       \fi is matched with {$\yy_i$}.
    }\label{fig:prefs}
  \end{figure}

  \paragraph{Initial matching~$M_0$.} It is defined as follows~(also see the agents marked in boxes in \cref{fig:prefs}).
  \begin{enumerate}[(i)]
    \item For each $v_i\in V$, define $M_0({\vvv_i})\coloneqq {\yy_i}$ and $M_0({\xx_i})={\ww_i}$.
    \item For each $e_j\in E$, define $M_0({\ff_j})\coloneqq {\ee_j}$ and $M_0({\pp_j})={\qq_j}$.
    \item For each ${\kk} \in [h]$, define $M_0(t_{\kk})\coloneqq b_{\kk}$ and $M(a_{\kk})\coloneqq s_{\kk}$.
    \item For each ${\kk}\in [\binom{h}{2}]$, define $M_0(c_{{\kk}}) \coloneqq z_{{\kk}}$ and $M_0(r_{{\kk}})\coloneqq d_{{\kk}}$.
  \end{enumerate}

  This completes the construction for the reduction, which can clearly be conducted in polynomial time.
  Before we continue with the correctness proof, we observe the following two technical properties regarding reachable matchings and stable matchings, respectively.
  \begin{claim}\label{claim:reachable-matching}
    For each two matchings~$N$ and $M$ such that $M$ is reachable from $N$ the following holds.
    \begin{enumerate}[(1)]
      \item\label{reachable-notsk} For each vertex-selector-agent~$s_k\in S$, if $N(s_k)\neq a_k$, then $M(s_k)\neq a_k$.
      \item\label{reachable-ck} For each edge-selector-agent~$d_{\kk}\in D$, if $N(d_{\kk}) = c_{\kk}$, then $M(d_{\kk}) = c_{\kk}$.
      \item\label{reachable-e} For each edge-agent~${\ff_j}\in F$, if $N({\ff_j})\neq {\ee_j}$, then $M({\ff_j})\neq {\ee_j}$.
    \end{enumerate}
  \end{claim}

  \iflong \begin{proof} \else \begin{proof}[Proof sketch.] \fi
    \renewcommand{\qedsymbol}{(of \cref{claim:reachable-matching})~\cqed}
    \ifshort
    We only prove Statement~\eqref{reachable-notsk} since the other proofs are of similar spirit.
    \fi
    Let $(\rho_0,\rho_1,\ldots,\rho_{\ell-1})$ denote a witness for $M$ to be reachable from $N$,
    and let $L=(N_0=N,N_{1},\ldots,N_{\ell}=M)$ be the corresponding sequence of matchings with
    $N_i=\dopm(N_{i-1},\rho_{i-1})$, $i\in [\ell]$.
   \iflong

   \noindent\textbf{Statement~\eqref{reachable-notsk}:} \fi
    Consider an arbitrary vertex-selector-agent~$s_k\in S$ such that $N_0(s_k)\neq a_k$.
    Suppose, for the sake of contradiction, that $N_{\ell}(s_k) = a_k$.
    This means that there exist two consecutive matchings in the sequence~$L$ where the partner of $s_k$ changes from some agent other than~$a_k$ to~$a_k$.
    Let $N_{\alpha-1}$ and $N_{\alpha}$, $\alpha \in [\ell]$, be two consecutive matchings witnessing this, i.e., 
    $N_{\alpha-1}(s_k) \neq a_k$ while $N_{\alpha}(s_k) = a_k$.
    Since $N_{\alpha} = \dopm(N_{\alpha-1}, \rho_{\alpha-1})$,
    by the definition of \dop{s}, it follows that $\rho_{\alpha-1}=\{s_k,a_k\}$ or $\rho_{\alpha-1}=\{N_{\alpha-1}(s_k),N_{\alpha-1}(a_k)\}$.
    Since $\rho_{\alpha-1}$ is blocking~$N_{\alpha-1}$ but $a_k$ is the least preferred agent of $s_k$,
    we infer that $\rho_{\alpha-1}=\{N_{\alpha-1}(s_k),N_{\alpha-1}(a_k)\}$.
    Clearly, by the acceptable agents of~$a_k$, we have that $N_{\alpha-1}(a_k) \in \{b_k\}\cup Y$. %
    Since $a_k$ is the most preferred agent of $b_k$ it follows that $N_{\alpha-1}(a_k) \neq b_k$ as otherwise $\{N_{\alpha-1}(s_k), N_{\alpha-1}(a_k)\}$ would not be blocking~$N_{\alpha-1}$.
    Hence, $N_{\alpha-1}(a_k)={\yy_i}$ for some~$i\in [n]$.
    By the acceptable agents of $s_k$, we have that $N_{\alpha-1}(s_k)\in \{t_k\}\cup V$.
    However, no agent from $V$ prefers~$y_i$ to~$s_k$ and $t_k$ is not acceptable to~$\yy_i$,
    a contradiction for~$\{N_{\alpha-1}(s_k), N_{\alpha-1}(a_k)\}$ being a blocking pair of $N_{\alpha-1}$.
    \iflong

    \noindent\textbf{Statement~\eqref{reachable-ck}:}
    The reasoning is analogous to the one for Statement~\eqref{reachable-notsk}.
    Consider an arbitrary edge-selector-agent~$d_k\in D$ such that $N_0(d_k) = c_k$.
    Suppose, for the sake of contradiction, that $N_{\ell}(d_k) \neq c_k$.
    This means that there exist two consecutive matchings in the sequence~$L$ where the partner of $d_k$ changes from agent~$c_k$ to someone other than $c_k$.
    Let $N_{\alpha-1}$ and $N_{\alpha}$, $\alpha \in [\ell]$, be such two consecutive matchings witnessing the change, i.e., 
    $N_{\alpha-1}(d_k) = c_k$ while $N_{\alpha}(d_k) \neq c_k$.
    Since $N_{\alpha} = \dopm(N_{\alpha-1}, \rho_{\alpha-1})$,
    by the definition of \dop{s}, it follows that $\rho_{\alpha-1}=\{d_k,N_{\alpha}(d_k)\}$ or $\rho_{\alpha-1}=\{c_k,N_{\alpha}(c_k)\}$.
    Since $\rho_{\alpha-1}$ is blocking~$N_{\alpha-1}$ but $c_k$ is the most preferred agent of $d_k$,
    we infer that $\rho_{\alpha-1}=\{c_k,N_{\alpha}(c_k)\}$ with $N_{\alpha}(c_k) \in \{z_k\}\cup E$.
    Since $c_k$ is the least preferred agent of $z_k$ it follows that $N_{\alpha}(c_k) \neq z_k$ as otherwise $\rho_{\alpha-1}=\{c_k, N_{\alpha}(c_k)\}$ would not be blocking $N_{\alpha-1}$.
    This means that $N_{\alpha}(c_k) = {\ee_j}$ for some~$j\in [m]$.
    Since $\{c_k, {\ee_j}\}$ is blocking~$N_{\alpha-1}$, meaning that $\ee_j$ prefers $c_k$ to~$N_{\alpha}({d_k})$,
    by the preferences of $\ee_j$ it follows that $N_{\alpha}(d_k)\in C\setminus \{c_k\}$,
    a contradiction since $d_k$ does not find any agent from $C\setminus \{c_k\}$ acceptable.

    \noindent \textbf{Statement~\eqref{reachable-e}:}
    The reasoning is similar to the one for Statement~\eqref{reachable-notsk}.
    Consider an arbitrary edge-agent~${\ff_j}\in F$ such that $N_0({\ff_j})\neq {\ee_j}$.
    Suppose, for the sake of contradiction, that $N_{\ell}({\ff_j}) = {\ee_j}$.
    This means that there exist two consecutive matchings in the sequence~$L$ where the partner of ${\ff_j}$ changes from some agent other than~${\ee_j}$ to~${\ee_j}$.
    Let $N_{\alpha-1}$ and $N_{\alpha}$, $\alpha \in [\ell]$, be two consecutive matchings witnessing this, i.e., 
    $N_{\alpha-1}({\ff_j}) \neq {\ee_j}$ while $N_{\alpha}({\ff_j}) = {\ee_j}$.
    Since $N_{\alpha} = \dopm(N_{\alpha-1}, \rho_{\alpha-1})$,
    by the definition of \dop{s}, it follows that $\rho_{\alpha-1}=\{{\ff_j},{\ee_j}\}$ or $\rho_{\alpha-1}=\{N_{\alpha-1}({\ff_j}),N_{\alpha-1}({\ee_j})\}$.
    Since $\rho_{\alpha-1}$ is blocking~$N_{\alpha-1}$ but ${\ee_j}$ is the least preferred agent of~${\ff_j}$,
    we infer that $\rho_{\alpha-1}=\{N_{\alpha-1}({\ff_j}),N_{\alpha-1}({\ee_j})\}$.
    Clearly, $N_{\alpha-1}({\ff_j})\in \{{\qq_j}\}\cup D$.
    However, $N_{\alpha-1}(\ff_j)$ cannot be $\qq_j$ because except $\ff_j$ no other agent prefers~$\qq_j$ to~$\ee_j$.
    Hence, $N_{\alpha-1}({\ff_j})=d_k$ for some $k\in [\binom{h}{2}]$.
    This implies that $N_{\alpha-1}({\ee_j})=r_k$ because except~$\ff_j$ only $r_k$ prefers $d_k$ to~$\ee_j$.
    However, $d_k$ does not prefer $r_k$ to $\ff_j$, a contradiction to $\rho_{\alpha-1}$ being a blocking pair of $N_{\alpha-1}$.
    \fi
  \end{proof}

  The above claim specifies how the partners of some agents may change in a reachable matching, 
  while the following claim specifies how a stable matching would look like.

  \begin{claim}\label{claim:SM-prop}
    Every stable matching~$M$ of our constructed instance must satisfy the following.
    \begin{enumerate}[(1)]
      \item\label{SM-prop:SE} For each~${\kk}\in [\binom{h}{2}]$, either ``$M(c_{{\kk}})=d_{{\kk}}$ and $M(r_{{\kk}})=z_{{\kk}}$'' or ``$M(c_{{\kk}})=z_{{\kk}}$ and $M(r_{{\kk}})=d_{{\kk}}$''.
      \item\label{SM-prop:E} For each~$j\in [m]$, either ``$M({\ff_j})={\ee_j}$ and $M({\pp_j}) = {\qq_j}$'' or ``$M({\ff_j})={\qq_j}$ and $M({\pp_j})={\ee_j}$''.
    \end{enumerate}
  \end{claim}
  \iflong
  \begin{proof}\renewcommand{\qedsymbol}{(of \cref{claim:SM-prop})~\cqed}
    Let $M$ denote an arbitrary stable matching.

  \noindent\textbf{Statement~\eqref{SM-prop:SE}:} %
  We distinguish between two cases, where ${\kk}\in [\binom{h}{2}]$.
     \begin{description}
      \item[Case 1:] $M({c_{{\kk}}})={z_{{\kk}}}$. By the preferences of agents~${z_{{\kk}}}$ and $r_{\kk'}$,
      it must hold that $M({r_{{\kk}}})={d_{{\kk}}}$ as otherwise $\{{r_{{\kk}}},z_{{\kk}}\}$ would be blocking~$M$.
      \item[Case 2:] $M(c_{{\kk}})\neq z_{{\kk}}$. By the preferences of agent~$c_{{\kk}}$ it must hold that $M(z_{{\kk}})=r_{{\kk}}$ as otherwise $\{c_{{\kk}}, z_{{\kk}}\}$ would be blocking~$M$.
      Moreover, since $c_{{\kk}}$ is the most preferred agent of~$d_{{\kk}}$, it must hold that $M(c_{{\kk}})\in \{d_{{\kk}}\}\cup E$ as otherwise $c_{\kk'}$ and $d_{\kk'}$ would form a blocking pair.
      If $M(c_{{\kk}})\in E$, say $M(c_{{\kk}})=e_{j}$ for some $j\in [m]$, then $\{p_j,e_j\}$ would be blocking~$M$.
      Hence, $M(c_{{\kk}})=d_{{\kk}}$.
  \end{description}
  
    \noindent \textbf{Statement~\eqref{SM-prop:E}:}
    This can be shown analogously, distinguishing between two cases, where $j\in [m]$:
    \begin{description}
      \item[Case 1:] $M({\ff_j})={\qq_j}$. By the preferences of~${\qq_j}$ and ${\pp_j}$ we immediately have that $M({\pp_j})={\ee_j}$ as otherwise $\{{\pp_j}, {\qq_j}\}$ would be blocking~$M$.
      \item[Case 2:] $M({\ff_j})\neq {\qq_j}$.
      Since $\qq_j$ is the most preferred agent of~$\ff_j$,
      it follows that $M({\qq_j})\in \{{\pp_j}\}\cup C$.
      By Statement~\eqref{SM-prop:SE}, $M({\qq_j})={\pp_j}$.
      Moreover, since $M({\ff_j})\notin D$ (see Statement~\eqref{SM-prop:SE})
      and since $\ff_j$ is the most preferred agent of ${\ee_j}$,
      it must hold that~$M({\ff_j})={\ee_j}$.      \hfill \qedhere
    \end{description} 
  \end{proof}
  \fi
  Now, we are ready to show that graph~$G$ admits a clique with $h$ vertices if and only if the constructed instance has a stable matching which is reachable from~$M_0$.
  For the ``only if'' direction, assume that $G$ admits a clique with $h$ vertices and let $V'$ denote such a vertex subset~$V'$ with $|V'|=h$ such that $G[V']$ is a complete subgraph.
  To ease notation, let $V'=\{v_{i_1}, v_{i_2}, \ldots, v_{i_{h}}\}$ such that $i_1<i_2<\cdots < i_{h}$, and let
  $E'=\{e_{j_1}, e_{j_2}, \ldots, e_{j_{\binom{h}{2}}}\}$ denote the edge set in the induced complete subgraph~$G[V']$ such that $j_1 < j_2 < \cdots < j_{\binom{h}{2}}$.

  \noindent We perform several stages of \dop{s} in order to obtain a stable matching.
  Briefly put, we will perform $4\binom{h}{2}$ \dop{s} involving edge gadgets that correspond to the ``clique edges''.
  As already discussed, these \dop{s} make $\binom{h}{2}$ edge-agents unhappy since each of them prefers to be with either of its incident vertex-agents.
  However, it is not possible to resolve this since the respective partners are not unacceptable to each other. 
  Hence, we then perform $4h$ \dop{s} involving vertex gadgets that correspond to the ``clique vertices''.
  In this way, no blocking pair remains.
  \begin{enumerate}[(1)]
    \item For each ${\kk}\in [h]$, we perform four \dop{s} as follows:
    Define
    \iflong
    \begin{enumerate}[(i)]
      \item $M_{4{\kk}-3}=\dopm(M_{4{\kk}-4}, \{{\vvv_{i_\kk}}, {s_{\kk}}\})$,
      \item $M_{4{\kk}-2}=\dopm(M_{4{\kk}-3}, \{{\xx_{i_\kk}}, b_{\kk}\})$,
      \item  $M_{4{\kk}-1}=\dopm(M_{4{\kk}-2}, \{t_{\kk}, {s_{\kk}}\})$,  and
      \item $M_{4{\kk}} = $ $\dopm(M_{4{\kk}-1}$, $\{a_{\kk}, {b_\kk}\})$.
    \end{enumerate}
    \else
     (i)  $M_{4{\kk}-3}=\dopm(M_{4{\kk}-4}, \{{\vvv_{i_\kk}}, {s_{\kk}}\})$,
     (ii) $M_{4{\kk}-2}=\dopm(M_{4{\kk}-3}, \{{\xx_{i_\kk}}, b_{\kk}\})$,
     (iii)  $M_{4{\kk}-1}=\dopm(M_{4{\kk}-2}, \{t_{\kk}, {s_{\kk}}\})$,  and
     (iv) $M_{4{\kk}} = $ $\dopm(M_{4{\kk}-1}$, $\{a_{\kk}, {b_\kk}\})$.
     \fi
     The four \dop{s} are depicted in \cref{fig:dops-vertex-gadget}.
     \iflong

     \fi
     Note that since $\{{\vvv_{i_{\kk}}}, {s_{i_\kk}}\}$ is blocking~$M_{4{\kk}-4}$,
    and since $M_{4{\kk}-4}({\vvv_{i_\kk}})=M_0({\vvv_{i_\kk}})={\yy_{i_\kk}}$ and $M_{4{\kk}-4}({s_{\kk}})={{a_{\kk}}}$
    can be matched together,
    we obtain that $M_{4{\kk}-3}$ is indeed a matching.
    Thus, $M_{4{\kk}-3}$ can be obtained from $M_{4{\kk}-4}$ by performing a \dop{} by $\{\vvv_{i_\kk}, {s_{\kk}}\}$.
     \iflong

     \fi
    Analogously, since $\{{\xx_{i_{\kk}}}, {b_{\kk}}\}$ is blocking~$M_{4{\kk}-3}$,
    and since $M_{4{\kk}-3}({\xx_{i_\kk}})=M_0({\xx_{i_\kk}})={\ww_{i_\kk}}$ and $M_{4{\kk}-3}({b_{\kk}})={{t_{\kk}}}$
    can be matched together,
    we obtain that $M_{4{\kk}-2}$ is indeed a matching.
    Thus, $M_{4{\kk}-2}$ can be obtained from $M_{4{\kk}-3}$ by performing a \dop{} by $\{{\xx_{i_\kk}}, {b_{\kk}}\}$.
     \iflong

     \fi
    Note that these first two \dop{s} are independent and can be interchanged.

    After that, $M_{4{\kk}-2}({t_{\kk}})={\ww_{i_{\kk}}}$ while $M_{4{\kk}-2}({s_{\kk}})={\vvv_{i_{\kk}}}$, implying that both $\{t_\kk, s_\kk\}$ and $\{{\vvv_{i_{\kk}}}, {\ww_{i_{\kk}}}\}$ are blocking~$M_{4{\kk}-2}$.
    Hence, $M_{4{\kk}-1}$ can be obtained from $M_{4{\kk}-2}$ by  performing a \dop{} by for instance $\{t_{\kk}, s_\kk\}$.
     \iflong

     \fi
    Analogously, $M_{4{\kk}-1}(a_{\kk})={\yy_{i_{\kk}}}$ while $M_{4{\kk}-1}(b_{{\kk}})={\xx_{i_{\kk}}}$, implying that
    $\{a_{\kk}, b_{{\kk}}\}$ and $\{{\xx_{i_{\kk}}}, {\yy_{i_{\kk}}}\}$ are both blocking~$M_{4{\kk}-1}$.
    Thus, $M_{4{\kk}}$ can be obtained from $M_{4{\kk}-1}$ by  performing a \dop{} for instance by $\{a_{\kk}, b_\kk\}$.  
     \iflong

     \fi
    After the above $4h$ \dop{s}, no blocking pairs of $M_ {4h}$ involve any agent from $S\cup B \cup \{{\vvv_{i}}, {\xx_{i}} \mid v_i \in V'\} \cup \{{\ww_i}, {\yy_i} \mid v_i \in V\setminus V'\}$ since each of these agents already obtains her most preferred partner. 
  
    \begin{figure}[t!]
      \begin{tikzpicture}
      \node (case) {};
      \node[boy, below = 0pt of case] (vi) {};  
      \node[girl, below = \ydist of vi] (yi) {};  
      \node[boy, below = \ydist of yi] (aell) {};  
      \node[girl, below = \ydist of aell] (sell) {};

      \node[girl, right = \xdist of vi] (wi) {};  
      \node[boy, right = \xdist of yi] (xi) {};  
      \node[girl, right = \xdist of aell] (bell) {};  
      \node[boy, right = \xdist of sell] (tell) {};

      \gettikzxy{(vi)}{\vx}{\vy};
      \gettikzxy{(bell)}{\yx}{\yyy};
      \gettikzxy{(xi)}{\xxx}{\xy};
      \gettikzxy{(wi)}{\wx}{\wy};
      \node at (\vx*0.5+\wx*0.5, \vy+4ex) {$M_{4{\kk}-4}:$};
      \node at (\yx+7ex, \xy*0.5+\yyy*0.5) {$\leadsto$};
      
      \foreach \i / \n / \r in {vi/{\vvv_{i_{\kk}}}/left,xi/{\xx_{i_{\kk}}}/right,
        tell/{t_{{\kk}}}/right, aell/{a_{{\kk}}}/left,
        wi/{\ww_{i_{\kk}}}/right, yi/{\yy_{i_{\kk}}}/left,
        sell/{s_{{\kk}}}/left, bell/{b_{{\kk}}}/right}{
        \node[\r = 0pt of \i] (n\i) {$\n$};
      }
      
      \foreach \s / \t in {vi/yi, xi/wi, tell/bell, aell/sell} {
        \path%
        (\s) edge[thick] (\t);
      }
  
    \end{tikzpicture}~
    \begin{tikzpicture}
      \node (case) {};
      \node[boy, below = 0pt of case] (vi) {};  
      \node[girl, below = \ydist of vi] (yi) {};  
      \node[boy, below = \ydist of yi] (aell) {};  
      \node[girl, below = \ydist of aell] (sell) {};

      \node[girl, right = \xdist of vi] (wi) {};  
      \node[boy, right = \xdist of yi] (xi) {};  
      \node[girl, right = \xdist of aell] (bell) {};  
      \node[boy, right = \xdist of sell] (tell) {};

      \gettikzxy{(vi)}{\vx}{\vy};
      \gettikzxy{(bell)}{\yx}{\yyy};
      \gettikzxy{(xi)}{\xxx}{\xy};
      \gettikzxy{(wi)}{\wx}{\wy};
      \node at (\vx*0.5+\wx*0.5, \vy+4ex) {$M_{4{\kk}-3}:$};
      \node at (\yx+7ex, \xy*0.5+\yyy*0.5) {$\leadsto$};
      
      \foreach \i / \n / \r in {vi/{\vvv_{i_{\kk}}}/left,xi/{\xx_{i_{\kk}}}/right,
        tell/{t_{{\kk}}}/right, aell/{a_{{\kk}}}/left,
        wi/{\ww_{i_{\kk}}}/right, yi/{\yy_{i_{\kk}}}/left,
        sell/{s_{{\kk}}}/left, bell/{b_{{\kk}}}/right}{
        \node[\r = 0pt of \i] (n\i) {$\n$};
      }

      \foreach \s / \t / \ang in {vi/sell/20, xi/wi/0, tell/bell/0, aell/yi/0} {
        \path%
        (\s) edge[thick, bend left = \ang] (\t);
      }
    \end{tikzpicture}~
    \begin{tikzpicture}
      \node (case) {};
      \node[boy, below = 0pt of case] (vi) {};  
      \node[girl, below = \ydist of vi] (yi) {};  
      \node[boy, below = \ydist of yi] (aell) {};  
      \node[girl, below = \ydist of aell] (sell) {};

      \node[girl, right = \xdist of vi] (wi) {};  
      \node[boy, right = \xdist of yi] (xi) {};  
      \node[girl, right = \xdist of aell] (bell) {};  
      \node[boy, right = \xdist of sell] (tell) {};

      \gettikzxy{(vi)}{\vx}{\vy};
      \gettikzxy{(bell)}{\yx}{\yyy};
      \gettikzxy{(xi)}{\xxx}{\xy};
      \gettikzxy{(wi)}{\wx}{\wy};
      \node at (\vx*0.5+\wx*0.5, \vy+4ex) {$M_{4{\kk}-2}:$};
      \node at (\yx+7ex, \xy*0.5+\yyy*0.5) {$\leadsto$};
      
      \foreach \i / \n / \r in {vi/{\vvv_{i_{\kk}}}/left,xi/{\xx_{i_{\kk}}}/right,
        tell/{t_{{\kk}}}/right, aell/{a_{{\kk}}}/left,
        wi/{\ww_{i_{\kk}}}/right, yi/{\yy_{i_{\kk}}}/left,
        sell/{s_{{\kk}}}/left, bell/{b_{{\kk}}}/right}{
        \node[\r = 0pt of \i] (n\i) {$\n$};
      }

      \foreach \s / \t / \ang in {vi/sell/20, xi/bell/0, tell/wi/20, aell/yi/0} {
        \path%
        (\s) edge[thick, bend left = \ang] (\t);
      } 
    \end{tikzpicture}~
    \begin{tikzpicture}
      \node (case) {};
      \node[boy, below = 0pt of case] (vi) {};  
      \node[girl, below = \ydist of vi] (yi) {};  
      \node[boy, below = \ydist of yi] (aell) {};  
      \node[girl, below = \ydist of aell] (sell) {};

      \node[girl, right = \xdist of vi] (wi) {};  
      \node[boy, right = \xdist of yi] (xi) {};  
      \node[girl, right = \xdist of aell] (bell) {};  
      \node[boy, right = \xdist of sell] (tell) {};

      \gettikzxy{(vi)}{\vx}{\vy};
      \gettikzxy{(bell)}{\yx}{\yyy};
      \gettikzxy{(xi)}{\xxx}{\xy};
      \gettikzxy{(wi)}{\wx}{\wy};
      \node at (\vx*0.5+\wx*0.5, \vy+4ex) {$M_{4{\kk}-1}:$};
      \node at (\yx+7ex, \xy*0.5+\yyy*0.5) {$\leadsto$};
      
      \foreach \i / \n / \r in {vi/{\vvv_{i_{\kk}}}/left,xi/{\xx_{i_{\kk}}}/right,
        tell/{t_{{\kk}}}/right, aell/{a_{{\kk}}}/left,
        wi/{\ww_{i_{\kk}}}/right, yi/{\yy_{i_{\kk}}}/left,
        sell/{s_{{\kk}}}/left, bell/{b_{{\kk}}}/right}{
        \node[\r = 0pt of \i] (n\i) {$\n$};
      }

      \foreach \s / \t / \ang in {vi/wi/0, xi/bell/0, tell/sell/0, aell/yi/0} {
        \path%
        (\s) edge[thick, bend left = \ang] (\t);
      }
    \end{tikzpicture}~
   \begin{tikzpicture}
     \node (case) {};
       \node[boy, below = 0pt of case] (vi) {};  
      \node[girl, below = \ydist of vi] (yi) {};  
      \node[boy, below = \ydist of yi] (aell) {};  
      \node[girl, below = \ydist of aell] (sell) {};

      \node[girl, right = \xdist of vi] (wi) {};  
      \node[boy, right = \xdist of yi] (xi) {};  
      \node[girl, right = \xdist of aell] (bell) {};  
      \node[boy, right = \xdist of sell] (tell) {};  

      \gettikzxy{(vi)}{\vx}{\vy};
      \gettikzxy{(bell)}{\yx}{\yyy};
      \gettikzxy{(xi)}{\xxx}{\xy};
      \gettikzxy{(wi)}{\wx}{\wy};
      \node at (\vx*0.5+\wx*0.5, \vy+4ex) {$M_{4{\kk}}:$};
      
      \foreach \i / \n / \r in {vi/{\vvv_{i_{\kk}}}/left,xi/{\xx_{i_{\kk}}}/right,
        tell/{t_{{\kk}}}/right, aell/{a_{{\kk}}}/left,
        wi/{\ww_{i_{\kk}}}/right, yi/{\yy_{i_{\kk}}}/left,
        sell/{s_{{\kk}}}/left, bell/{b_{{\kk}}}/right}{
        \node[\r = 0pt of \i] (n\i) {$\n$};
      }  
      
      \foreach \s / \t / \ang in {vi/wi/0, xi/yi/0, tell/sell/0, aell/bell/0} {
        \path%
        (\s) edge[thick, bend left = \ang] (\t);
      }   
    \end{tikzpicture} 
    \caption{The four \dop{s} for involving the vertex-agents~$\{{\vvv_{i_k}}, {\xx_{i_k}}, {\ww_{i_k}}, {\yy_{i_k}}\}$ used in the proof of \cref{thm:divor-incomplete-noties-nphard}.}\label{fig:dops-vertex-gadget} 
    \end{figure} 
    \item    Let $N_0=M_{4{h}}$. 
    For each ${\kk}'\in [\binom{h}{2}]$, we again perform four \dop{s} as follows:
    define
    \iflong
    \begin{enumerate}[(i)]
      \item $N_{4{\kk}'-3}=\dopm(N_{4{\kk}'-4}, \{{\ff_{j_{\kk'}}}, d_{{\kk}'}\})$,
      \item  $N_{4{\kk}'-2}=\dopm(N_{4{\kk}'-3}, \{r_{{\kk}'}, z_{{\kk}'}\})$,
      \item  $N_{4{\kk}'-1}=\dopm(N_{4{\kk}-2}, \{{\pp_{j_{{\kk}'}}}, {\ee_{j_{{\kk}'}}}\})$,
      and
      \item $N_{4{\kk}'}=\dopm(N_{4{\kk}'-1}, \{c_{{\kk}'}, d_{{\kk}'}\})$.
    \end{enumerate}
    \else
     (i) $N_{4{\kk}'-3}=\dopm(N_{4{\kk}'-4}, \{{\ff_{j_{\kk'}}}, d_{{\kk}'}\})$,
     (ii) $N_{4{\kk}'-2}=\dopm(N_{4{\kk}'-3}, \{r_{{\kk}'}, z_{{\kk}'}\})$,
     (iii)  $N_{4{\kk}'-1}=\dopm(N_{4{\kk}-2}, \{{\pp_{j_{{\kk}'}}}, {\ee_{j_{{\kk}'}}}\})$,
      and
     (iv) $N_{4{\kk}'}=\dopm(N_{4{\kk}'-1}, \{c_{{\kk}'}, d_{{\kk}'}\})$.
      \fi
    The \dop{s} are depicted in \cref{fig:dops-edge-gadget}.
    One can verify that for each matching~$M$ and pair~$\rho$ in the above operation, $M$ is blocked by~$\rho$ and the partners of the agents in~$\rho$ (under~$M$) can be matched to each other.
    \iflong

    \fi
    After the above \dop{s}, it is straightforward to see that no blocking pairs involve any agent from~$D\cup Z\cup \{{\ff_{j}}, {\pp_{j}} \mid e_j\in E'\}\cup \{{\ee_j},{\qq_j}\mid e_j\notin E'\}$ as every such agent already obtains her most preferred agent. 
  \end{enumerate}
    \begin{figure}[t!]
    \begin{tikzpicture}
      \node (case) {};
      \node[boy, right = 2ex of case] (pi) {};  
      \node[girl, below = \ydist of pi] (qi) {};  
      \node[boy, below = \ydist of qi] (cell) {};  
      \node[girl, below = \ydist of cell] (zell) {};  

      \node[girl, right = \xdist of pi] (ei) {};  
      \node[boy, right = \xdist of qi] (fi) {};  
      \node[girl, right = \xdist of cell] (dell) {};  
      \node[boy, right = \xdist of zell] (rell) {};  

      \foreach \i / \n / \r in {fi/{\ff_{j_{{\kk}'}}}/right,pi/{\pp_{j_{{\kk}'}}}/left,
        cell/{c_{{\kk}'}}/left, rell/{r_{{\kk}'}}/right,
        ei/{\ee_{j_{{\kk}'}}}/right, qi/{\qq_{j_{{\kk}'}}}/left,
        dell/{d_{{\kk}'}}/right, zell/{z_{{\kk}'}}/left}{
        \node[\r = 0pt of \i] (n\i) {$\n$};
      }

      \gettikzxy{(pi)}{\vx}{\vy};
      
      \gettikzxy{(ei)}{\wx}{\wy};

      \gettikzxy{(fi)}{\fx}{\fy}
      \gettikzxy{(dell)}{\dx}{\dy}
      \node at (\fx+7ex, \fy*0.5+\dy*0.5) {$\leadsto$};
      
      \node at (\vx*0.5+\wx*0.5, \vy+4ex) {$N_{4{\kk}'-4}:$};
      
      \foreach \s / \t in {fi/ei, pi/qi, cell/zell, rell/dell} {
        \path%
        (\s) edge[thick] (\t);
      }
  
    \end{tikzpicture}~
    \begin{tikzpicture}
      \node (case) {};
      \node[boy, right = 2ex of case] (pi) {};  
      \node[girl, below = \ydist of pi] (qi) {};  
      \node[boy, below = \ydist of qi] (cell) {};  
      \node[girl, below = \ydist of cell] (zell) {};  

      \node[girl, right = \xdist of pi] (ei) {};  
      \node[boy, right = \xdist of qi] (fi) {};  
      \node[girl, right = \xdist of cell] (dell) {};  
      \node[boy, right = \xdist of zell] (rell) {};  

      \foreach \i / \n / \r in {fi/{\ff_{j_{{\kk}'}}}/right,pi/{\pp_{j_{{\kk}'}}}/left,
        cell/{c_{{\kk}'}}/left, rell/{r_{{\kk}'}}/right,
        ei/{\ee_{j_{{\kk}'}}}/right, qi/{\qq_{j_{{\kk}'}}}/left,
        dell/{d_{{\kk}'}}/right, zell/{z_{{\kk}'}}/left}{
        \node[\r = 0pt of \i] (n\i) {$\n$};
      }

      \gettikzxy{(pi)}{\vx}{\vy};
      
      \gettikzxy{(ei)}{\wx}{\wy};

      \gettikzxy{(fi)}{\fx}{\fy}
      \gettikzxy{(dell)}{\dx}{\dy}
      \node at (\fx+7ex, \fy*0.5+\dy*0.5) {$\leadsto$};
      
      \node at (\vx*0.5+\wx*0.5, \vy+4ex) {$N_{4{\kk}'-3}:$};
      
      \foreach \s / \t / \ang in {fi/dell/0, pi/qi/0, cell/zell/0, rell/ei/-20} {
        \path%
        (\s) edge[thick, bend right=\ang] (\t);
      }
  
    \end{tikzpicture}~
    \begin{tikzpicture}
      \node (case) {};
       \node[boy, right = 2ex of case] (pi) {};  
      \node[girl, below = \ydist of pi] (qi) {};  
      \node[boy, below = \ydist of qi] (cell) {};  
      \node[girl, below = \ydist of cell] (zell) {};  

      \node[girl, right = \xdist of pi] (ei) {};  
      \node[boy, right = \xdist of qi] (fi) {};  
      \node[girl, right = \xdist of cell] (dell) {};  
      \node[boy, right = \xdist of zell] (rell) {};  

      \foreach \i / \n / \r in {fi/{\ff_{j_{{\kk}'}}}/right,pi/{\pp_{j_{{\kk}'}}}/left,
        cell/{c_{{\kk}'}}/left, rell/{r_{{\kk}'}}/right,
        ei/{\ee_{j_{{\kk}'}}}/right, qi/{\qq_{j_{{\kk}'}}}/left,
        dell/{d_{{\kk}'}}/right, zell/{z_{{\kk}'}}/left}{
        \node[\r = 0pt of \i] (n\i) {$\n$};
      }

      \gettikzxy{(pi)}{\vx}{\vy};
      
      \gettikzxy{(ei)}{\wx}{\wy};

      \gettikzxy{(fi)}{\fx}{\fy}
      \gettikzxy{(dell)}{\dx}{\dy}
      \node at (\fx+7ex, \fy*0.5+\dy*0.5) {$\leadsto$};
      
      \node at (\vx*0.5+\wx*0.5, \vy+4ex) {$N_{4{\kk}'-2}:$};
      
      \foreach \s / \t / \ang in {fi/dell/0, pi/qi/0, cell/ei/0, rell/zell/0} {
        \path%
        (\s) edge[thick, bend right=\ang] (\t);
      }
    \end{tikzpicture}~
    \begin{tikzpicture}
      \node (case) {};
      \node[boy, right = 2ex of case] (pi) {};  
      \node[girl, below = \ydist of pi] (qi) {};  
      \node[boy, below = \ydist of qi] (cell) {};  
      \node[girl, below = \ydist of cell] (zell) {};  

      \node[girl, right = \xdist of pi] (ei) {};  
      \node[boy, right = \xdist of qi] (fi) {};  
      \node[girl, right = \xdist of cell] (dell) {};  
      \node[boy, right = \xdist of zell] (rell) {};  

      \foreach \i / \n / \r in {fi/{\ff_{j_{{\kk}'}}}/right,pi/{\pp_{j_{{\kk}'}}}/left,
        cell/{c_{{\kk}'}}/left, rell/{r_{{\kk}'}}/right,
        ei/{\ee_{j_{{\kk}'}}}/right, qi/{\qq_{j_{{\kk}'}}}/left,
        dell/{d_{{\kk}'}}/right, zell/{z_{{\kk}'}}/left}{
        \node[\r = 0pt of \i] (n\i) {$\n$};
      }

      \gettikzxy{(pi)}{\vx}{\vy};
      
      \gettikzxy{(ei)}{\wx}{\wy};

      \gettikzxy{(fi)}{\fx}{\fy}
      \gettikzxy{(dell)}{\dx}{\dy}
      \node at (\fx+7ex, \fy*0.5+\dy*0.5) {$\leadsto$};
      
      \node at (\vx*0.5+\wx*0.5, \vy+4ex) {$N_{4{\kk}'-1}:$};
      
      \foreach \s / \t / \ang in {fi/dell/0, pi/ei/0, cell/qi/0, rell/zell/0} {
        \path%
        (\s) edge[thick, bend right=\ang] (\t);
      }
    \end{tikzpicture}~
    \begin{tikzpicture}
      \node (case) {};
      \node[boy, right = 2ex of case] (pi) {};  
      \node[girl, below = \ydist of pi] (qi) {};  
      \node[boy, below = \ydist of qi] (cell) {};  
      \node[girl, below = \ydist of cell] (zell) {};  

      \node[girl, right = \xdist of pi] (ei) {};  
      \node[boy, right = \xdist of qi] (fi) {};  
      \node[girl, right = \xdist of cell] (dell) {};  
      \node[boy, right = \xdist of zell] (rell) {};  

      \foreach \i / \n / \r in {fi/{\ff_{j_{{\kk}'}}}/right,pi/{\pp_{j_{{\kk}'}}}/left,
        cell/{c_{{\kk}'}}/left, rell/{r_{{\kk}'}}/right,
        ei/{\ee_{j_{{\kk}'}}}/right, qi/{\qq_{j_{{\kk}'}}}/left,
        dell/{d_{{\kk}'}}/right, zell/{z_{{\kk}'}}/left}{
        \node[\r = 0pt of \i] (n\i) {$\n$};
      }

      \gettikzxy{(pi)}{\vx}{\vy};
      
      \gettikzxy{(ei)}{\wx}{\wy};

      \gettikzxy{(fi)}{\fx}{\fy}
      \gettikzxy{(dell)}{\dx}{\dy}
      
      \node at (\vx*0.5+\wx*0.5, \vy+4ex) {$N_{4{\kk}'}:$};
      
      \foreach \s / \t / \ang in {fi/qi/0, pi/ei/0, cell/dell/0, rell/zell/0} {
        \path%
        (\s) edge[thick, bend right=\ang] (\t);
      }
    \end{tikzpicture}
    \caption{The four \dop{s} for involving the edge-agents~$\{{\ff_{i_k}}, {\pp_{i_k}}, {\ee_{i_k}}, {\qq_{i_k}}\}$ used in the proof of \cref{thm:divor-incomplete-noties-nphard}.}\label{fig:dops-edge-gadget}
  \end{figure}
  Observe that in total, we have performed~$4h+4\binom{h}{2}$~\dop{s}. 
  It remains to show that $N_{4{\binom{h}{2}}}$ is stable.
  To ease notation, let $M\coloneqq N_{4{\binom{h}{2}}}$.
  Since we are in the bipartite case, to show stability, we will show that no blocking pair involves an agent from $\hat{W}$.
  More precisely, we show the following:
  \begin{itemize}[--]
    \item Consider an arbitrary~$i\in [n]$. Clearly, if $v_i\in V\setminus V'$, then no blocking pair involves ${\ww_i}$ or ${\yy_i}$ since they both have their most preferred agents.
    If $v_i\in V'$, then by our definition, $M({\ww_i})={\vvv_i}$ while $M({\yy_i})={\xx_i}$.
    Clearly, ${\ww_i}$ cannot be involved in any blocking pair as she only prefers ${\xx_i}$ to her partner but ${\xx_i}$ already obtains her most preferred agent.
    Analogously, ${\yy_i}$ is also not involved in any blocking pair as she only prefers ${\vvv_i}$ to her partner but ${\vvv_i}$ already obtains her most preferred agent.

    \item By our construction, every agent from $S\cup B$ already obtained her most preferred agent.
    \item Consider an arbitrary~$j\in [m]$.
    Again, clearly, if $e_j\in E\setminus E'$,
    then no blocking pair involves ${\ee_j}$ or ${\qq_j}$ since they both obtain their most preferred agents.
    
    It remains to consider the case when $e_j\in E'$.
    In this case, by our definition of the sequence of \dop{s}, $M({\ee_j})={\pp_j}$ while $M({\qq_j})={\ff_j}$.
    Clearly, ${\qq_j}$ is not involved in any blocking pair as each agent from~$C$ prefers her partner (which is someone from~$D$) to~${\qq_j}$, and ${\pp_j}$ already obtains her most preferred agent.

    Neither does ${\ee_j}$ form with any agent from $R\cup \{{\ff_j}\}\cup C$ a blocking pair since every agent from~$R$ prefers her partner (which is someone from~$Z$) to~${\ee_j}$,
    agent~$\ff_j$ already obtains her most preferred agent,
    and ${\ee_j}$ prefers her partner~$p_j$ to all agents from~$C$.

    It remains to consider the ``incident'' vertex-agents~${\vvv_i}, {\vvv_{i'}}$ with $e_j=\{v_i,v_{i'}\}$.
    Since $e_j\in E'$, meaning that $v_i,v_{i'}\in V'$, by our construction, we have that $M({\vvv_i})={\ww_i}$ and $M({\vvv_{i'}})={\ww_{i'}}$.
    This means that both ${\vvv_i}$ and ${\vvv_{i'}}$ already obtain their most preferred agents and will not form with ${\ee_j}$ a blocking pair.

  \end{itemize}
  Since no agent from $\hat{W}$ is involved in any blocking pair, we infer that $M$ is stable.

  For the ``if'' part of the correctness proof, assume that there exists a stable matching, denoted as~$M_\ell$, which is reachable from~$M_0$.
  Let $L'=(\rho_0,\rho_1,\ldots,\rho_{\ell-1})$ be a witness for~$M_\ell$ to be reachable from~$M_0$.
  Before we show how to construct an $h$-vertex clique for~$G=(V,E)$, we explain the intuitive idea. 
  Observe that each vertex-selector-agent from~$S$ (resp.\ $B$) will help a unique vertex-agent~${\vvv_i}\in V$ (resp.\ $\xx_i\in X$) in reaching her most preferred agent, namely~${\ww_i}$ (resp.\ ${\yy_i}$)
  and each edge-selector-agent from~$D$ will help a unique edge-agent~${\ff_j}\in F$ in reaching her most preferred agent, namely~${\qq_j}$.
  Hence, by \cref{claim:SM-prop}\eqref{SM-prop:E}, agent~${\ee_j}$ will need to be matched to~${\pp_j}$.
  By the preferences of the vertex-agents from~$V$ and the edge-agents from~$E$, this means that the two ``incident'' vertex-agents of~${\ee_j}$ must \emph{not} be matched with their initial partners as otherwise they will form with $\ee_j$ a blocking pair.
  To achieve this, they will need the ``help'' of the vertex-selector-agents. 
  By the number of vertex-selector-agents, there must be exactly $h$ such agents which correspond to a clique of size~$h$.

  We formalize the above idea through the following technical properties for the sequence~$L'$ which will guide us to select a clique solution for~$G$.
  
  \begin{claim}\label{claim:reachable-SM}
    For each~$i\in [\ell]$, define $M_{i}\coloneqq \dopm(M_{i-1}, \rho_{i-1})$.
    Then, the following holds.
    \begin{enumerate}[(1)]

       \item\label{reachable-SM-V}
       For each vertex-agent~${\vvv_i} \in V$ with $M_{\ell}({\vvv_i})\neq {\yy_i}$
       there exist a selector-agent~$s_k\in S$ and an index~$\alpha\in [\ell]$ such that $M_{\alpha-1}({\vvv_i}) = {\yy_i}$, $M_{\alpha-1}({a_k}) = s_k$, 
       $M_{\alpha}({\vvv_i}) = {s_k}$, and $M_{\alpha}({a_k}) = {\yy_i}$.

        \item \label{reachable-SM-D0-unique} For each~$d_k\in D$ it holds that $M_{\ell}(d_{k})=c_{k}$.
      \item\label{reachable-SM-D1}
      For each selector-agent~$d_k\in D$,
      there exist an edge-agent~${\ff_j}\in F$ and an index~$\alpha\in [\ell]$ such that
      $M_{\alpha-1}(d_k) = r_k$, $M_{\alpha-1}({\ee_j}) = {\ff_j}$,
      $M_{\alpha}(d_k) = {\ff_j}$, and $M_{\alpha}({\ee_j}) = {r_k}$.

    \end{enumerate}
  \end{claim}
  \iflong
  \begin{proof} \renewcommand{\qedsymbol}{(of \cref{claim:reachable-SM})~\cqed}
    Let $M_1,M_2,\ldots,M_{\ell}$ be as defined and define $L\coloneqq (M_0,M_1,\ldots,M_{\ell})$.

    \noindent \textbf{Statement~\eqref{reachable-SM-V}:}
    Consider an arbitrary vertex-agent~${\vvv_i}\in V$ with $M_{\ell}({\vvv_i})\neq {\yy_i}$.
    By the definition of the initial matching~$M_0$, there must be two consecutive matchings~$M_{\alpha-1}$ and $M_{\alpha}$ in~$L$, $\alpha\in [\ell]$,
    such that $M_{\alpha-1}({\vvv_i}) = {\yy_i}$
    while $M_{\alpha}({\vvv_i})\neq {\yy_i}$.
    Since $M_{\alpha}=\dopm(M_{\alpha-1}, \rho_{\alpha-1})$, by the definition of \dop{s},
    $\rho_{\alpha-1}=\{{\vvv_i}, M_{\alpha}({\vvv_i})\}$ or $\rho_{\alpha-1}=\{{\yy_i},
    M_{\alpha}({\yy_i})\}$.
    Since $\vvv_i$ is the most preferred agent of~$\yy_i$,
    we infer that $\rho_{\alpha-1}=\{{\vvv_i}, M_{\alpha}({\vvv_i})\}$. %
    By the acceptable partners of~${\yy_i}$ it follows that %
    $M_{\alpha}({\yy_i}) \in \{{\xx_i}\}\cup A$.
    Since no agent, except~$\yy_i$, prefers~${\vvv_i}$ to~${\xx_i}$, we further infer that
    $M_{\alpha}({\yy_i}) = a_k$ for some $k\in [h]$.
    Since~$s_k$ is the only acceptable agent of~$\vvv_i$ (except~$\yy_i$) who prefers~$\vvv_i$ to~$a_k$,
    it follows that $M_{\alpha}({\vvv_i})=s_k$; note that $\{{\vvv_i}, M_{\alpha}({\vvv_i})\}$ is blocking~$M_{\alpha-1}$.
    This implies that $M_{\alpha-1}(a_k)=s_k$.
    Summarizing, we have found such a vertex selector-agent~$s_k\in S$ and an index~$\alpha$ for the statement.
    
    \noindent\textbf{Statement~\eqref{reachable-SM-D0-unique}:}
    Suppose, for the sake of contradiction, that there exists an edge-selector-agent~$d_k\in D$ with $M_{\ell}(d_k)\neq c_k$.
    By \cref{claim:SM-prop}\eqref{SM-prop:SE}, it follows that $M_{\ell}(d_k) = r_k$.
    Then, by the preferences of the $F$-agents and by \cref{claim:SM-prop}\eqref{SM-prop:E},
    for each $j\in [m]$, it must hold that $M_{\ell}({\ff_j}) = {\qq_j}$ as otherwise
     $M_{\ell}({\ff_j}) = {\ee_j}$ so~$\{{\ff_j}, d_k\}$ would be blocking~$M_\ell$.
    Consider an arbitrary edge-agent~${\ff_j}\in F$.
    Since $M_0({\ff_j})={\ee_j}\neq {\qq_j}$,
    there exist two consecutive matchings in~$L$ where the partner of~${\ff_j}$ changes from someone other than~${\qq_j}$ to~${\qq_j}$.
    Let $M_{\alpha-1}$ and $M_{\alpha}$, $\alpha\in [\ell]$, be two consecutive matchings witnessing this, i.e., $M_{\alpha-1}({\ff_j})\neq {\qq_j}$ and $M_{\alpha}({\ff_j})={\qq_j}$.
    Since $M_{\alpha}=\dopm(M_{\alpha-1}, \rho_{\alpha-1})$, by the definition of \dop{s},
    it follows that
    $\rho_{\alpha-1}=\{{\ff_j}, {\qq_j}\}$ or $\rho_{\alpha-1}=\{M_{\alpha-1}({\ff_j}), M_{\alpha-1}({\qq_j})\}$.
    Since ${\ff_j}$ is the least preferred agent of ${\qq_j}$,
    we infer that $\rho_{\alpha-1}=\{M_{\alpha-1}({\ff_j}), M_{\alpha-1}({\qq_j})\}$.
    Since ${\ff_j}$ is the most preferred agent of~${\ee_j}$,
    we also infer that $M_{\alpha-1}({\ff_j})\neq {\ee_j}$.
    By the acceptable partners of~${\ff_j}$ it follows that $M_{\alpha-1}({\ff_j})=d_{k}$ for some $k\in [\binom{h}{2}]$.
    Observe that besides~${\ff_j}$ only~$c_k$ finds both~${\qq_j}$ and $d_k$ acceptable.
    This implies that $M_{\alpha-1}({\qq_j})=c_k$ and $M_{\alpha}(d_k)=c_k$.
    By \cref{claim:reachable-matching}\eqref{reachable-ck}, $d_{k}$ and $c_{k}$ remain matched to each other in $(M_{\alpha}, M_{\alpha+1}, \ldots, M_{\ell})$, a contradiction to $M_{\ell}(d_k) \neq c_k$.

    \noindent \textbf{Statement~\eqref{reachable-SM-D1}:}
    The reasoning is analogous to the one for Statement~\eqref{reachable-SM-V}.
    Consider an arbitrary edge-selector-agent~$d_k\in D$.
    By Statement~\eqref{reachable-SM-D0-unique} and since $M_0(d_k)=r_k$,
    there must be two consecutive matchings~$M_{\alpha-1}$ and $M_{\alpha}$ in~$L$, $\alpha\in [\ell]$,
    such that $M_{\alpha-1}(d_k) = r_k$ while $M_{\alpha}(d_k)\neq r_k$.
    Since $M_{\alpha}=\dopm(M_{\alpha-1}, \rho_{\alpha-1})$,
    it follows that
    $\rho_{\alpha-1}=\{d_k, M_{\alpha}(d_k)\}$ or $\rho_{\alpha-1}=\{r_k, M_{\alpha}(r_k)\}$.
    Since $d_k$ is $r_k$'s most preferred agent, 
    we infer that $\rho_{\alpha-1}=\{d_k, M_{\alpha}(d_k)\}$.
    By the acceptable agents of~$r_k$, it follows that $M_{\alpha}(r_k)\in \{z_k\}\cup E$.
    Since $M_{\alpha}(d_k)$ and $M_{alpha}(r_k)$ are matched in $M_{\alpha-1}$,
    we infer that $M_{\alpha}(r_k)\neq z_k$ since no agent, except~$d_k$, prefers $d_k$ to~$z_k$ so
    that $\{d_k, M_{\alpha}(d_k)\}$ cannot be blocking~$M_{\alpha-1}$.
    This means that $M_{\alpha}(r_k) = {\ee_j}$ for some $j\in [m]$.
    Since, except~$r_k$, agent~${\ff_j}$ is the only agent who prefers~$d_k$ to $\ee_j$,
    we infer that $M_{\alpha}(d_k)={\ff_j}$.
    Summarizing, we have found such an edge-agent~${\ff_j}$ and an index~$\alpha\in [\ell]$ for the statement.
  \end{proof}
  \fi
  Now, we claim that $G$ admits a clique of size~$h$.
  For technical reasons, we define the following
  edge and vertex subsets~$E'\coloneqq \{e_j \in E \mid M_{\alpha}({\ff_j}) = d_k \text{ for some } k \in [\binom{h}{2}] \text{ and } \alpha \in [\ell] \}$
  and $V'\coloneqq \{v_i,v_{i'} \mid \{v_i,v_{i'}\} = e_j \text{ for some } e_j \in E'\}$.
  We claim that $V'$ is a clique of size~$h$.
  To show this, we only need to show that
  \begin{align}
    |E'| & \ge \binom{h}{2}, \text{ and }\label{if-dir:E'-card}\\
    |V'| & \le h;\label{if-dir:V'-card}
  \end{align}
  note that for any edge subset~$H$ of cardinality~$\binom{h}{2}$ the number of endpoints of the edges in~$H$ is at least~$h$ and it is $h$ if and only if these endpoints form a clique.

  To show Inequality~\eqref{if-dir:E'-card}, for each edge-selector-agent~$d_k\in D$ let ${\ff_{j_k}}$ and $\alpha_k$ denote an edge-agent and an index mentioned according to~\cref{claim:reachable-SM}\eqref{reachable-SM-D1}.
  Define the following sequence~$F'\coloneqq ({\ff_{j_1}}, {\ff_{j_2}}, \ldots, {\ff_{j_k}})$. %
  We claim that no two agents~${\ff_{j_k}}, {\ff_{j_{k'}}}\in F'$ are the same.
  Without loss of generality, assume that $\alpha_k < \alpha_{k'}$; note that $\alpha_k\neq \alpha_{k'}$.
  This means that $M_{\alpha_{k'}-1}$ is reachable from~$M_{\alpha_k}$.
  By our definition of ${\ff_{j_k}}$ and $\alpha_k$ we have that $M_{\alpha_{k}}({\ff_{j_{k}}})\neq {\ee_{j_k}}$.
  Since $M_{\alpha_{k'}-1}$ is reachable from $M_{\alpha_k}$, by \cref{claim:reachable-matching}\eqref{reachable-e} (applied for $\ff_{j_k}$ and matchings~$M_{\alpha_{k}}$ and $M_{\alpha_{k'}-1}$), it follows that $M_{\alpha_{k'}-1}({\ff_{j_{k}}}) \neq {\ee_{j_k}}$.
  Hence $j_k \neq j_{k'}$ since $M_{\alpha_{k'}-1}({\ff_{j_{k'}}}) = {\ee_{j_{k'}}}$.
  Summarizing, since $|D|=\binom{h}{2}$, there must be at least $\binom{h}{2}$ distinct edge-agents~$\ff_{j_{k}}$.
  By the definition of $E'$ we have that $|E'|\ge \binom{h}{2}$.

  Next, we show Inequality~\eqref{if-dir:V'-card}.
  Consider an arbitrary vertex~$v_i \in V'$.
  By the definition of $V'$, there exists an incident edge~$e_j\in E'$ with $v_i \in e_j$
  such that $M_{\alpha}({\ff_j}) = d_k$ for some~$k\in [\binom{h}{2}]$ and some~$\alpha\in [\ell]$.
  By \cref{claim:reachable-matching}\eqref{reachable-e}, $M({\ee_j})\neq {\ff_j}$ since $M$ is reachable from~$M_{\alpha}$.
  Since $M$ is stable, by \cref{claim:SM-prop}\eqref{SM-prop:E}, $M({\ee_j}) = {\pp_j}$.
  By the preferences of~${\vvv_i}$ and since $v_i\in e_j$, it follows that
  $M({\vvv_i})\neq {\yy_i}$ as otherwise ${\vvv_i}$ and ${\ee_j}$ would form a blocking pair of~$M$. 
  Observe that this holds for every vertex~$v_i \in V'$.
  By \cref{claim:reachable-SM}\eqref{reachable-SM-V}, let $s_{k_i}$ and $\alpha_i$ denote the vertex-selector-agent and index with $M_{\alpha_i-1}({\vvv_i})={\yy_i}$,
  $M_{\alpha_i-1}({a_{k_i}})={s_{k_i}}$,
  $M_{\alpha_i}({\vvv_i})={s_{k_i}}$, and $M_{\alpha_i}({a_{k_i}})={\yy_i}$. 
  Now, observe that if we can show that no two such vertex-selector-agents are the same, then we obtain that $|V'|\le h$ since $|S|=h$.

  Consider two vertex-agents~${\vvv_i},{\vvv_j}\in V'$, together with the just defined vertex-selector-agents~$s_{k_i}$ and $s_{k_{j}}$ and the corresponding indices~$\alpha_i$ and $\alpha_j$.
  In particular, we have that $M_{\alpha_i-1}(s_{k_i}) = a_{k_i}$ and $M_{\alpha_i}(s_{k_i}) = {\vvv_i}$.
  Without loss of generality, assume that $\alpha_i < \alpha_{j}$; note that $\alpha_i\neq \alpha_{j}$.
  This means that $M_{\alpha_j-1}$ is reachable from~$M_{\alpha_i}$.
  Since $M_{\alpha_i}(s_{k_i}) \neq a_{k_i}$, by \cref{claim:reachable-matching}\eqref{reachable-notsk} (applied for $s_{k_i}$ and matchings~$M_{\alpha_i}$ and $M_{\alpha_j-1}$), it follows that $M_{\alpha_j-1}({s_{k_i}}) \neq {a_{k_i}}$.
  We further infer that $k_j\neq k_i$ because otherwise by  \cref{claim:reachable-SM}\eqref{reachable-SM-V} (applied for ${\vvv_j}$) we must have that
  $M_{\alpha_{j}-1}({s_{k_j}})=M_{\alpha_{j}-1}({s_{k_i}})\neq a_{k_i} = a_{k_j}$, a contradiction to the definition of $s_{k_j}$.
  Summarizing, since $|S|=h$, there can be at most $h$~vertex-agents from~$V'$ whose partners are changed to the agents from~$W$, i.e., $|V'|\le h$.  
\end{proof}

\section{Parameterizations}\label{sec:param}
In this section, we consider the parameterized complexity of \divorcesms{}.
The parameters that we are interested in are the maximum number~$\kappa$ of allowed \dop{s} and the maximum length~$d$ of the preference lists.

\subsection{Max.\ number~$\kappa$ of \dop{s} allowed}
We observe that if the maximum number~$k$ of \dop{s} allowed is a constant, then we can solve \divorcesms{} in polynomial time by simply guessing the blocking pair in each \dop{} operation.

\begin{lemma}\label{cor:dop-xp}
  \divorcesms{} can be solved in $O(n^{2\kappa+2}\cdot \kappa^2)$~time, where $\kappa$ denotes the number of allowed \dop{s}.
\end{lemma}

\iflong \begin{proof}
  Let $I=(U,W,(\succ_x)_{x\in U\cup W}, M_0)$ be an instance of \divorcesms{} with $|U|=|W|=n$.
  Assume that $M_0$ can reach a stable matching, say~$M$, via~$\kappa'$ \dop{s} with $0\le \kappa' \le \kappa$ and let $L=(\rho_0,\rho_1,\ldots,\rho_{\kappa'})$ be a witness for $M$ to be reachable from~$M_0$.
  Since each of the blocking pairs in $L$ consists of two agents, there are at most $n^2$ different blocking pairs which we can simply guess.

  For each~$\kappa'$ with $0\le \kappa' \le \kappa$,
  we go through every possible sequence of blocking pairs of length~$\kappa'$ and check whether after performing the \dop{s} by the sequence we reach a stable matching.
  Since there are at most $\kappa \cdot n^{2\kappa}$ possible sequences, by \cref{lem:check-witness},
  we can check all of them in $O(n^{2\kappa+2}\cdot \kappa^2)$~time, as desired.
\end{proof}
\fi
The running time given in \cref{cor:dop-xp} cannot be improved substantially due to the parameterized intractability result as given in \cref{thm:dop-w1hard}.

\iflong \begin{proof}[Proof of \cref{thm:dop-w1hard}]
  To show W[1]-hardness, we use the same reduction as given for \cref{thm:divor-incomplete-noties-nphard} and set the number of allowed \dop{s} to~$\kappa\coloneqq 4h+4\binom{h}{2}$.
  Note that in the proof for \cref{thm:divor-incomplete-noties-nphard},
  the NP-complete \textsc{Clique} problem, from which we reduce to show NP-hardness,
  is W[1]-hard with respect to the solution parameter~``the size of the clique~$h$''.
  Moreover, in the ``only if'' direction, we actually showed that if there exists a clique of size~$h$, then we can reach a stable matching from the initial matching~$M_0$, using $4h+\binom{h}{2}$ \dop{s}.
  If we set the number of allowed \dop{s} to $\kappa\coloneqq 4h+\binom{h}{2}$, then the polynomial-time reduction given in the proof of \cref{thm:divor-incomplete-noties-nphard} is also a parameterized reduction  regarding the parameter~$\kappa$.
  The W[1]-hardness result follows since \textsc{Clique} is W[1]-hard with respect to~$h$.
\end{proof}
\fi

\subsection{Max.\ length~$d$ of the preference lists}

Although the proof of \cref{thm:divor-incomplete-noties-nphard} produces a \divorcesms{} instance, where the length of a preference list may be unbounded, in the following, we show that for preference lists of constant length, the problem remains NP-hard~(\cref{thm:NP-hard-constant-length}).

\iflong
\begin{proof}[Proof of \cref{thm:NP-hard-constant-length}]
  \else
  \begin{proof}[Proof sketch of \cref{thm:NP-hard-constant-length}]
    \fi
  The hardness reduction is quite similar to the one for \cref{thm:divor-incomplete-noties-nphard}.
  The main difficulty is to reduce the length of the preference lists.
  To achieve this, we will instead reduce from a restricted variant of \textsc{3SAT} which allows us to treat each variable (resp.\ each clause) ``independently'' in the sense that we can use a variable-selector-agent (resp.\ a clause-selector-agent) for each variable (resp.\ each clause) which guides us to select a truth value for each variable (resp.\ a truth-setting literal for each clause).
  The preferences of the literal-agents and the clause-agents ensure that each corresponding clause is satisfied by at least one literal.
  The restricted variant of \textsc{3SAT} is called \textsc{R3SAT}~\cite[p.~259]{GJ79} and has the property that each literal appears at most twice, guaranteeing that the length of the constructed preference lists is a constant.
  Formally, \textsc{R3SAT} has as input a set~$V=\{v_1,v_2,\ldots,v_n\}$ of variables and a set~$E=\{e_1,e_2,\ldots, e_m\}$ of clauses over~$V$ with at most three literals per clause such that each literal appears at most twice, and asks whether there exists a satisfying truth assignment for~$E$.
  Note that we select the symbols for the variables and the clauses to largely match the gadgets that we constructed in the proof for \cref{thm:divor-incomplete-noties-nphard}.
  We will adopt the edge-gadgets used in \cref{thm:divor-incomplete-noties-nphard}.
   
  Let $I=(V,E)$ be an instance of \textsc{R3SAT} with $V=\{v_1,v_2,\ldots, v_n\}$ and $E=\{e_1,e_2,\ldots, e_m\}$.
  For each clause~$e_j$, let $|e_j|$ denote the number of literals appearing in $e_j$.
  
  Our \divorcesms{} instance consists of two disjoint sets of agents, $\hat{U}$ and $\hat{W}$,
  with $\hat{U} \coloneqq V \uplus \overline{V}
  \uplus X \uplus \overline{X} \uplus T \uplus A \uplus F\uplus P \uplus C \uplus R$
  and $\hat{W} \coloneqq W \uplus \overline{W} \uplus Y \uplus \overline{Y} \uplus S \uplus B \uplus {\hat{E}} \uplus Q \uplus D \uplus Z$, where

  \noindent
  \begin{tabular}{@{}c@{\;}llll}
  -- &  $V\coloneqq \{{\vvv_i}\mid v_i\in V\}$, & $\overline{V}\coloneqq \{{\nvv_i}\mid v_i \in V\}$,
    & $X \coloneqq \{{\xx_i} \mid i\in [n]\}$, & $\overline{X} \coloneqq \{{\nxx_i} \mid i\in [n]\}$,\\
   &  $W \coloneqq \{{\ww_i} \mid i\in [n]\}$, & $\overline{W} \coloneqq \{{\nww_i} \mid i\in [n]\}$,
    & $Y\coloneqq \{{\yy_i} \mid i \in [n]\}$, & $\overline{Y}\coloneqq \{{\nyy_i} \mid i \in [n]\}$,\\
    --  & \multicolumn{2}{@{}l}{$\hat{E}\coloneqq \{{\ee_j^{\kk}} \mid j \in [m]\wedge \kk \in [|e_j|]\}$,} &  \multicolumn{2}{l}{$F \coloneqq \{{\ff^{\kk}_j} \mid j \in [m] \wedge \kk \in [|e_j|]\}$,} \\
     & \multicolumn{2}{@{}l}{$P\coloneqq \{{\pp^{\kk}_j} \mid j\in [m] \wedge \kk \in [|e_j|]\}$,} & \multicolumn{2}{l}{$Q\coloneqq \{{\qq^{\kk}_j} \mid j\in [m] \wedge \kk \in [|e_j|]\}$,}\\
  -- &  $S \coloneqq \{s_i \mid i \in [n]\}$, & $T\coloneqq \{t_i \mid i \in [n]\}$, & $A\coloneqq \{a_{i} \mid {i} \in [n]\}$, & $B\coloneqq \{b_{i} \mid {i} \in [n]\}$,\\
     & $C\coloneqq \{c_{j} \mid {j} \in [m]\}$, & $D\coloneqq \{d_{j} \mid {j} \in [m]\}$, & $R\coloneqq \{r_{j} \mid {j} \in [m]\}$, & $Z\coloneqq \{z_{j} \mid {j} \in [m]\}$.
  \end{tabular}
  
\iflong \noindent  In words:
 \begin{itemize}[--]
   \item
    For each variable~$v_i\in V$ we introduce four \myemph{literal-agents}~${\vvv_i},{\xx_i},{\ww_i},{\yy_i}$ for the un-negated variable~$v_i$,
    and four further \myemph{literal-agents}~${\nvv_i},{\nxx_i},{\nww_i},{\nyy_i}$ for the negated variable~$\overline{v}_i$.
   \item  For each clause~$e_j\in E$ we introduce $4 |e_j|$ \myemph{clause-agents}~${\ff^i_j}, {\pp^i_j}, {\ee^i_j}, {\qq^i_j}$, $i\in [|e_j|]$.
    \item   For each~${\kk} \in [n]$, we also introduce four \myemph{variable-selector-agents}~$t_{\kk},s_{\kk},a_{\kk}, b_{\kk}$ who shall ``deviate'' with the literal-agents corresponding to the literal that will be set $\true$.
  \item  Finally, for each~${\kk}\in [m]$,
    we introduce four further \myemph{clause-selector-agents}~$c_{\kk},d_{\kk},r_{\kk},z_{\kk}$ who shall deviate with the clause-agents whose corresponding literal satisfies the clause. 
  \end{itemize}
    \fi
    Note that to enhance the connection between the variables and their corresponding un-negated literal-agents we use the same symbol~$v_i$ for both the variable and the corresponding literal-agent~${\vvv_i}$.
    In total, there are $6n+2\sum_{j\in [m]}|e_j|+2m\le 6n+8m$ agents on each side.
  \noindent  The preferences of the agents are depicted in \cref{fig:prefs-fromR3SAT}; we omit the ``preferring'' symbol~$\succ$ for brevity.
  \begin{figure}[t!]
    \begin{align*}
      \begin{array}{l@{\;}r@{\,}l@{}r@{\,}l}
        \forall i\in [n]\colon & {\vvv_i} \colon & \boxed{\ww_i} \mysucc b_i \mysucc [{\color{orange!70!black}E(v_i)}] \mysucc {\yy_i} \mysucc s_i,
        & {\ww_i} \colon & {\xx_i}\mysucc t_i  \mysucc \boxed{\vvv_i}, \\
        \forall i\in [n]\colon & {\xx_i} \colon & \boxed{\yy_i}\mysucc s_i \mysucc {\ww_i},
        & {\yy_i}\colon & {\vvv_i} \mysucc \boxed{\xx_i},\\
        \forall i\in [n]\colon & {\nvv_i} \colon & \boxed{\nww_i} \mysucc b_i \mysucc [{\color{orange!70!black}E(\overline{v}_i)}] \mysucc {\nyy_i} \mysucc s_i,
        & {\nww_i} \colon & {\nxx_i}\mysucc t_i  \mysucc \boxed{\nvv_i}, \\
        \forall i\in [n]\colon & {\nxx_i} \colon & \boxed{\nyy_i}\mysucc s_i \mysucc {\nww_i},
        & {\nyy_i}\colon & {\nvv_i} \mysucc \boxed{\nxx_i},\\[.6ex]
        \forall i\in [n]\colon & {t_{i}}\colon & s_{i} \mysucc {\ww_i} \mysucc {\nww_i} \mysucc \boxed{b_i},
        & s_{i} \colon & \boxed{a_{i}} \mysucc {\vvv_i}\mysucc {\nvv_i} \mysucc t_i \mysucc {\xx_i}
                         \mysucc {\nxx_i}, \\
         \forall i\in [n]\colon & {a_{i}}\colon & b_i \mysucc \boxed{s_{i}}, 
        & b_{i} \colon &  \boxed{t_{i}}  \mysucc a_i \mysucc {\vvv_i} \mysucc {\nvv_i}, \\[2ex]
       \forall j\in [m],  \forall \kk\in [|e_j|]\colon &{\ff^{\kk}_j} \colon & {\qq^{\kk}_j} \mysucc d_j \mysucc \boxed{\ee^{\kk}_j},
        & {\ee^{\kk}_j}\colon & \boxed{\ff^{\kk}_j}\mysucc {r_j} \mysucc {\vvv}({\ee^{\kk}_j})
                                \mysucc  {\pp^{\kk}_j} \mysucc c_j,\\
       \forall j\in [m],  \forall \kk\in [|e_j|]\colon  &      {\pp^{\kk}_j}\colon & {\ee^{\kk}_j} \mysucc \boxed{\qq^{\kk}_j},
        & {\qq^{\kk}_j}\colon & \boxed{\pp^{\kk}_j}  \mysucc c_j \mysucc {\ff^{\kk}_j},\\[1.2ex]
        \forall j\in [m],  \forall \kk\in [|e_j|]\colon    & {c_{j}}\colon & \boxed{z_{j}} \mysucc {\ee^1_j} \mysucc \cdots \mysucc {\ee^{|e_j|}_j} \mysucc
                        d_{j} \mysucc {\qq^1_j} \mysucc \cdots  \mysucc {\qq^{|e_j|}_j},~~~~~
        &  {d_{j}}\colon &  c_{j} \mysucc {\ff^1_j} \mysucc \cdots \mysucc {\ff^{|e_j|}_j} \mysucc \boxed{r_{j}},\\
       \forall j\in [m]\colon    &     {r_{j}} \colon& \boxed{d_{j}} \mysucc z_{j} \mysucc {\ee^1_j} \mysucc \cdots \mysucc {\ee^{|e_j|}_j},
         &   {z_{j}}\colon & r_{j}\mysucc \boxed{c_{j}}. \\
      \end{array}
    \end{align*}
    \caption{Preferences of the agents constructed in the proof of \cref{thm:NP-hard-constant-length}.
      Here, for each literal~$\lit_i\in V\cup \overline{V}$, the expression~$E(\lit_i)$ refers to the set of clause-agents corresponding to the clauses which include literal~$\lit_i$ and 
      $[E(\lit_i)]$ denotes an arbitrary but fixed order of the agents in~$E(\lit_i)$.
      \iflong Note that $|E(\lit_i)|\le 2$ since each literal appears in at most two clauses. \fi
      For each clause~$e_j$, we order the literals in~$e_j$, using an arbitrary but fixed order,
      and for each~$k\in [|e_j|]$, we use $v({\ee^k_j})$ to denote the literal-agent from~$V\cup \overline{V}$ which corresponds to the $k^{\text{th}}$ literal in clause~$e_j$.
      For instance, if $e_j=(v_2,\overline{v}_3,v_5)$, then
      $v({\ee^{1}_j})={\vvv_2}$, $v({\ee^{1}_j})={\nvv_3}$, and $v({\ee^{2}_j})={\vvv_5}$. 
      Since each clause contains at most three literals, the longest preference list created in the instance has length eight (see agent~$c_j$).
      In the initial matching~$M_0$, each agent is matched with the one marked in the box.
      For instance, for each $i\in [n]$, the un-negated literal agent~{$\vvv_i$} is matched with {$\ww_i$}.
    }\label{fig:prefs-fromR3SAT}
  \end{figure}

  \iflong  \paragraph{Initial matching~$M_0$.}It is defined as follows~(also see the agents marked in boxes in \cref{fig:prefs-fromR3SAT}).
  \begin{enumerate}[(i)]
    \item For each $i\in [n]$, define
    $M_0({\vvv_i})\coloneqq {\ww_i}$, $M_0({\xx_i})\coloneqq {\yy_i}$,
    $M_0({\nvv_i})\coloneqq {\nww_i}$, $M_0({\nxx_i})\coloneqq {\nyy_i}$,
    $M_0({t_i})\coloneqq {b_i}$, and $M_0({a_i})\coloneqq{s_i}$.
    \item For each $e_j\in E$ and each~$\kk \in [|e_j|]$,
    define $M_0({\ff^{\kk}_j})\coloneqq {\ee^{\kk}_j}$, $M_0({\pp^{\kk}_j})\coloneqq{\qq^{\kk}_j}$,
    $M_0({c_j})\coloneqq {z_j}$, $M_0(r_j)\coloneqq{d_j}$.
  \end{enumerate}
  \else
  The initial matching~$M_0$ is indicated via the boxed agents in \cref{fig:prefs-fromR3SAT}.
  \fi
  \iflong

  \fi
  This completes the construction for the reduction.
  One can verify that except the clause-selector agents from~$C$, each agent's preference list contains at most six agents. The agents in~$C$ may contain up to eight agents in his preference list.
  Hence, the length of each constructed preference list is bounded by eight.
  \ifshort
  The correctness proof is similar to the one for \cref{thm:divor-incomplete-noties-nphard}.
  Due to space constraints, we defer the correctness proof to the appendix.
  \fi
  \iflong

  Before we continue with the correctness proof, we observe the following two technical properties regarding reachable matchings and stable matching, respectively.
  \begin{claim}\label{claim:R3SAT-reachable-matching}
    For each two matchings~$N$ and $M$ such that $M$ is reachable from $N$ the following holds.
    \begin{enumerate}[(1)]
      \item\label{R3SAT-reachable-sk} For each variable-selector-agent~$t_i\in T$, if $N(t_i) = s_i$, then $M(t_i) = s_i$.
      \item\label{R3SAT-reachable-ck} For each clause-selector-agent~$d_{j}\in D$, if $N(d_{j}) = c_{j}$, then $M(d_{j}) = c_{j}$.
      \item\label{R3SAT-reachable-v} For each $i\in [n]$,
      if $N({\vvv_i}) \neq {\ww_i}$, then $M({\vvv_i}) \neq {\ww_i}$, and
      if $N({\nvv_i}) \neq {\nww_i}$, then $M({\nvv_i}) \neq {\nww_i}$.      
      \item\label{R3SAT-reachable-e} For each clause-agent~${\ff^{\kk}_j}\in F$, if $N({\ff^{\kk}_j})\neq {\ee^{\kk}_j}$, then $M({\ff^{\kk}_j})\neq {\ee^{\kk}_j}$.
    \end{enumerate}
  \end{claim}
  \iflong \begin{proof}
    \renewcommand{\qedsymbol}{(of \cref{claim:R3SAT-reachable-matching})~\cqed}
    The proofs for Statements~\eqref{R3SAT-reachable-sk} and \eqref{R3SAT-reachable-ck} are analogous to the one given for \cref{claim:reachable-matching}\eqref{reachable-ck}.
    The proofs for Statements~\eqref{R3SAT-reachable-v} and \eqref{R3SAT-reachable-e} are analogous to the one given for \cref{claim:reachable-matching}\eqref{reachable-e}.
    We repeat here for the sake of completeness.
    
    Let $(\rho_0,\rho_1,\ldots,\rho_{\ell-1})$ denote a witness for $M$ to be reachable from $N$
    and let $L=(N_0=N,N_{1},\ldots,N_{\ell}=M)$ be the corresponding sequence of matchings with
    $N_i=\dopm(N_{i-1},\rho_{i-1})$, $i\in [\ell]$.
    
    \noindent\textbf{Statement~\eqref{R3SAT-reachable-sk}:}
    Consider an arbitrary clause-selector-agent~$t_i\in T$ such that $N_0(t_i) = s_i$.
    Suppose, for the sake of contradiction, that $N_{\ell}(t_i) \neq s_i$.
    This means that there exist two consecutive matchings in the sequence~$L$ where the partner of $s_i$ changes from agent~$t_i$ to someone other than $t_i$.
    Let $N_{\alpha-1}$ and $N_{\alpha}$, $\alpha \in [\ell]$, be such two consecutive matchings witnessing the change, i.e., 
    $N_{\alpha-1}(t_i) = s_i$ while $N_{\alpha}(t_i) \neq s_i$.
    Since $N_{\alpha} = \dopm(N_{\alpha-1}, \rho_{\alpha-1})$,
    by the definition of \dop{s}, it follows that $\rho_{\alpha-1}=\{t_i,N_{\alpha}(t_i)\}$ or $\rho_{\alpha-1}=\{s_i,N_{\alpha}(s_i)\}$.
    Since $\rho_{\alpha-1}$ is blocking~$N_{\alpha-1}$ but $s_i$ is the most preferred agent of~$t_i$,
    we infer that $\rho_{\alpha-1}=\{s_i,N_{\alpha}(s_i)\}$ with $N_{\alpha}(s_i) \in \{a_i, {\vvv_i},
    {\nvv_i}\}$.
    However, every agent from~$ \{a_i, {\vvv_i}, {\nvv_i}\}$ regards~$s_i$ as the least preferred (acceptable) agent and will not form with $s_i$ a blocking pair, a contradiction.
    
    \noindent\textbf{Statement~\eqref{R3SAT-reachable-ck}:}
    Consider an arbitrary clause-selector-agent~$d_j\in D$ such that $N_0(d_j) = c_j$.
    Suppose, for the sake of contradiction, that $N_{\ell}(d_j) \neq c_j$.
    This means that there exist two consecutive matchings in the sequence~$L$ where the partner of $d_j$ changes from agent~$c_j$ to someone other than $c_j$.
    Let $N_{\alpha-1}$ and $N_{\alpha}$, $\alpha \in [\ell]$, be such two consecutive matchings witnessing the change, i.e., 
    $N_{\alpha-1}(d_j) = c_j$ while $N_{\alpha}(d_j) \neq c_j$.
    Since $N_{\alpha} = \dopm(N_{\alpha-1}, \rho_{\alpha-1})$,
    by the definition of \dop{s}, it follows that $\rho_{\alpha-1}=\{d_j,N_{\alpha}(d_j)\}$ or $\rho_{\alpha-1}=\{c_j,N_{\alpha}(c_j)\}$.
    Since $\rho_{\alpha-1}$ is blocking~$N_{\alpha-1}$ but $c_j$ is the most preferred agent of~$d_j$,
    we infer that $\rho_{\alpha-1}=\{c_j,N_{\alpha}(c_j)\}$ with $N_{\alpha}(c_j) \in \{z_j, {\ee^1_j}, \ldots, {\ee^{|e_j|}_j}\}$.
    However, every agent from~$ \{z_j, {\ee^1_j}, \ldots, {\ee^{|e_j|}_j}\}$ regards~$c_j$ as the least preferred (acceptable) agent and will not form with $c_j$ a blocking pair, a contradiction.    

    \noindent \textbf{Statement~\eqref{R3SAT-reachable-v}:}
    We only show the case with the un-negated literal~$v_i\in V$, the case with the negated literal~$\nvv_i$ works analogously.
    Consider an arbitrary literal-agent~${\vvv_i}\in V$ such that $N_0({\vvv_i})\neq {\ww_i}$.
    Suppose, for the sake of contradiction, that $N_{\ell}({\vvv_i}) = {\ww_i}$.
    This means that there exist two consecutive matchings in the sequence~$L$ where the partner of ${\vvv_i}$ changes from some agent other than~${\ww_i}$ to~${\ww_i}$.
    Let $N_{\alpha-1}$ and $N_{\alpha}$, $\alpha \in [\ell]$, be two consecutive matchings witnessing this, i.e., 
    $N_{\alpha-1}({\vvv_i}) \neq {\ww_i}$ while $N_{\alpha}({\vvv_i}) = {\ww_i}$.
    Since $N_{\alpha} = \dopm(N_{\alpha-1}, \rho_{\alpha-1})$,
    by the definition of \dop{s}, it follows that $\rho_{\alpha-1}=\{{\vvv_i},{\ww_i}\}$ or $\rho_{\alpha-1}=\{N_{\alpha-1}({\vvv_i}),N_{\alpha-1}({\ww_i})\}$.
    Since $\rho_{\alpha-1}$ is blocking~$N_{\alpha-1}$ but ${\vvv_i}$ is the least preferred agent of~${\ww_i}$,
    we infer that $\rho_{\alpha-1}=\{N_{\alpha-1}({\vvv_i}),N_{\alpha-1}({\ww_i})\}$.
    Clearly, $N_{\alpha-1}({\ww_i})\in \{{\xx_i}, t_i\}$.
    However, $N_{\alpha-1}({\ww_i})$ cannot be ${\xx_i}$ because except $\ww_i$ no other agent prefers~$\xx_i$ to~$\vvv_i$.
    Hence, $N_{\alpha-1}({\ww_i})=t_i$.
    This implies that $N_{\alpha-1}({\vvv_i})=b_i$ because except~$\ww_i$ only $b_i$ prefers $t_i$ to~$\vvv_i$.
    However, $t_i$ does not prefer $b_i$ to~${\ww_i}$, a contradiction to $\rho_{\alpha-1}$ being a blocking pair of $N_{\alpha-1}$.
    
    \noindent \textbf{Statement~\eqref{R3SAT-reachable-e}:}
    Consider an arbitrary clause-agent~${\ff^{\kk}_j}\in F$ such that $N_0({\ff^{\kk}_j})\neq {\ee^{\kk}_j}$.
    Suppose, for the sake of contradiction, that $N_{\ell}({\ff^{\kk}_j}) = {\ee^{\kk}_j}$.
    This means that there exist two consecutive matchings in the sequence~$L$ where the partner of ${\ff^{\kk}_j}$ changes from some agent other than~${\ee^{\kk}_j}$ to~${\ee^{\kk}_j}$.
    Let $N_{\alpha-1}$ and $N_{\alpha}$, $\alpha \in [\ell]$, be two consecutive matchings witnessing this, i.e., 
    $N_{\alpha-1}({\ff^{\kk}_j}) \neq {\ee^{\kk}_j}$ while $N_{\alpha}({\ff^{\kk}_j}) = {\ee^{\kk}_j}$.
    Since $N_{\alpha} = \dopm(N_{\alpha-1}, \rho_{\alpha-1})$,
    by the definition of \dop{s}, it follows that $\rho_{\alpha-1}=\{{\ff^{\kk}_j},{\ee^{\kk}_j}\}$ or $\rho_{\alpha-1}=\{N_{\alpha-1}({\ff^{\kk}_j}),N_{\alpha-1}({\ee^{\kk}_j})\}$.
    Since $\rho_{\alpha-1}$ is blocking~$N_{\alpha-1}$ but ${\ee^{\kk}_j}$ is the least preferred agent of~${\ff^{\kk}_j}$,
    we infer that $\rho_{\alpha-1}=\{N_{\alpha-1}({\ff^{\kk}_j}),N_{\alpha-1}({\ee^{\kk}_j})\}$.
    Clearly, $N_{\alpha-1}({\ff^{\kk}_j})\in \{{\qq^{\kk}_j}, {d_j}\}$.
    However, $N_{\alpha-1}(\ff^{\kk}_j)$ cannot be $\qq^{\kk}_j$ because except $\ff^{\kk}_j$ no other agent prefers~$\qq^{\kk}_j$ to~$\ee^{\kk}_j$.
    Hence, $N_{\alpha-1}({\ff^{\kk}_j})=d_j$.
    This implies that $N_{\alpha-1}({\ee^{\kk}_j})=r_j$ because except~$\ff^{\kk}_j$ only $r_j$ prefers $d_j$ to~$\ee^{\kk}_j$.
    However, $d_j$ does not prefer $r_j$ to~${\ff^{\kk}_j}$, a contradiction to $\rho_{\alpha-1}$ being a blocking pair of $N_{\alpha-1}$.
  \end{proof}
  \fi
  
  The above claim specifies how the partners of some agents may change in a reachable matching, 
  while the following claim specifies how a stable matching would look like.

  \begin{claim}\label{claim:R3SAT-SM-prop}
    Every stable matching~$M$ of our constructed instance must satisfy the following.
    \begin{enumerate}[(1)]
     \item\label{R3SAT-SM-prop:SV} For each~${i}\in [n]$, either ``$M(t_{i})=s_{i}$ and $M(a_{i})=b_{i}$'' or ``$M(t_{i})=b_{i}$ and $M(a_{i})=s_{i}$''.
      \item\label{R3SAT-SM-prop:SE} For each~$j \in [m]$, either ``$M(c_{j})=d_{j}$ and $M(r_{j})=z_{j}$'' or ``$M(c_{j})=z_{j}$ and $M(r_{j})=d_{j}$''. 
      \item\label{R3SAT-SM-prop:V} For each~$i\in [n]$, either ``$M({\vvv_i})={\ww_i}$ and $M({\xx_i}) = {\yy_i}$'' or ``$M({\vvv_i})={\yy_i}$ and $M({\xx_i})={\ww_i}$''.
      \item\label{R3SAT-SM-prop:negV} For each~$i\in [n]$, either ``$M({\nvv_i})={\nww_i}$ and $M({\nxx_i}) = {\nyy_i}$'' or ``$M({\nvv_i})={\nyy_i}$ and $M({\nxx_i})={\nww_i}$''.
      \item\label{R3SAT-SM-prop:E} For each~$j\in [m]$ and each~$\kk\in {[|e_j|]}$, either ``$M({\ff^{\kk}_i})={\ee^{\kk}_i}$ and $M({\pp^{\kk}_i}) = {\qq^{\kk}_i}$'' or ``$M({\ff^{\kk}_i})={\qq^{\kk}_i}$ and $M({\pp^{\kk}_i})={\ee^{\kk}_i}$''.
    \end{enumerate}
  \end{claim}
  \iflong
  \begin{proof}\renewcommand{\qedsymbol}{(of \cref{claim:R3SAT-SM-prop})~\cqed}
    Again, the proofs are analogous to the proof for~\cref{claim:SM-prop}
    Let $M$ denote an arbitrary stable matching.
    We only show \eqref{R3SAT-SM-prop:SV} and \eqref{R3SAT-SM-prop:negV} and omit the analogous proofs for the remaining statements.
    
    \noindent\textbf{Statement~\eqref{R3SAT-SM-prop:SV}:}
    We again distinguish between two cases, where $i\in [n]$.
     \begin{description}
      \item[Case 1:] $M(a_i)={s_i}$. By the preference of agent~${a_i}$,
      it must hold that $M({t_i})={b_i}$ as otherwise $\{{a_i},b_i\}$ would be blocking~$M$.
      \item[Case 2:] $M(a_i)\neq s_i$. By the preference of agent~$s_i$ it must hold that $M(a_i)=
      b_i$ as otherwise $\{a_i, s_i\}$ would be blocking~$M$.
      Moreover, since $s_i$ is the most preferred agent of~$t_i$,
      it must hold that $M(s_i)\in \{t_i,{\vvv_i}, {\nvv_i}\}$.
      If $M(s_i)\in \{{\vvv_i}, {\nvv_i}\}$, then $\{{\vvv_i}, {\yy_i}\}$ or $\{{\nvv_i}, {\nyy_i}\}$
      would be blocking~$M$.
      Hence, $M(s_{i})=t_i$.
    \end{description}

     \noindent \textbf{Statement~\eqref{R3SAT-SM-prop:negV}:}
    This can be shown analogously, distinguishing between two cases, where $i\in [n]$:
    \begin{description}
      \item[Case 1:] $M({\nxx_i})={\nww_i}$. By the preferences of~${\nxx_i}$ and $\nyy_i$,
      we immediately have that $M({\nyy_i})={\nvv_i}$ as otherwise $\{\nyy_i,\nvv_i\}$ would be blocking~$M$.
      \item[Case 2:] $M({\nxx_i})\neq {\nww_i}$.
      Since $\nww_i$ is the most preferred agent of~$\nvv_i$ and $M(t_i)\neq {\nww_i}$ (see Statement~\eqref{R3SAT-SM-prop:SV}),
      we have that $M(\nww_i) = {\nvv_i}$.
      Moreover, since $M({\nxx_i})\neq s_i$ (see Statement~\eqref{R3SAT-SM-prop:SE}) and since $\nxx_i$ is the most preferred agent of~$\nww_i$,
      it must hold that~$M({\nxx_i})={\nyy_i}$.\hfill \qedhere
    \end{description}
  \end{proof}
  \fi
  Now, we are ready to show that $I=(V,E)$ admits a satisfying assignment if and only if the constructed instance has a stable matching which is reachable from~$M_0$.
  For the ``only if'' direction, assume that $\sigma\coloneqq V\to \{\true,\false\}$ is a satisfying truth assignment for~$I$.
  For notational convenience, for each clause~$e_j\in E$ and each $k\in [|e_j|]$,
  let \myemph{$v({\ee^{k}_j})$} denote the literal-agent which corresponds to the $k^{\text{th}}$ literal in clause~$e_j$, and let \myemph{$k_j\in [|e_j|]$} denote an arbitrary but fixed index such that the $k_j^{\text{th}}$ literal in~$e_j$ is set $\true$ under~$\sigma$. 

  We perform two stages of \dop{s} in order to obtain a stable matching.
  \begin{enumerate}[(1)]
    \item For each $j\in [m]$, we perform four \dop{s} in the clause gadget for clause~$e_j$ as follows; recall that $k_j$ was defined as an index such that the $k_j^{\text{th}}$ literal in $e_j$ is set $\true$ under~$\sigma$:
    \begin{itemize}[--]
      \item $M_{4j-3}=\dopm(M_{4j-4}, \{{\ff^{k_j}_{j}}, d_{j}\})$,
      \item $M_{4j-2}=\dopm(M_{4j-3}, \{r_{j}, z_{j}\})$,
      \item $M_{4j-1}=\dopm(M_{4j-2}, \{{\pp^{k_j}_{j}}, {\ee^{k_j}_{j}}\})$,
      and
      \item $M_{4j}=\dopm(M_{4j-1}, \{c_{j}, d_{j}\})$.
    \end{itemize}
    The \dop{s} are depicted in \cref{fig:R3SAT-dops-edge-gadget}.
    We explain in the following why they are admissible.
    Since $\{{\ff^{k_j}_{j}}, {d_j}\}$ is blocking~$M_{4{j}-4}$,
    and since $M_{4{j}-4}({\ff^{k_j}_{j}})=M_0({\ff^{k_j}_{j}})={\ee^{k_j}_{j}}$ and $M_{4{j}-4}({d_{j}})={{r_j}}$ are acceptable to each other, 
    we obtain that $M_{4j-3}$ is a matching.
    Thus, $M_{4{j}-3}$ can be obtained from $M_{4{j}-4}$ by performing a \dop{} by $\{{\ff^{k_j}_{j}}, {d_j}\}$.

    Analogously, since $\{r_j, z_j\}$ is blocking~$M_{4{j}-3}$ (observe that $M_{4j-3}(r_j)={\ee^{k_j}_{j}}$),
    and since $M_{4{j}-3}(r_j)={\ee^{k_j}_{j}}$ and $M_{4{j}-3}(z_j)=M_0(z_j)={c_j}$
    are acceptable to each other,
    we obtain that $M_{4{j}-2}$ is indeed a matching.
    Thus, $M_{4{j}-2}$ can be obtained from $M_{4{j}-3}$ by performing a \dop{} by $\{r_j, z_j\}$.

    After that, $M_{4{j}-2}({\pp^{k_j}_{j}})=M_0({\pp^{k_j}_{j}})={\qq^{k_j}_{j}}$ while $M_{4{j}-2}({\ee^{k_j}_{j}})={c_j}$, implying that $\{{\pp^{k_j}_{j}}, {{\ee^{k_j}_{j}}}\}$ is blocking~$M_{4j-2}$. %
    Since ${\qq^{k_j}_{j}}$ and $c_j$ are acceptable to each other, %
     $M_{4{j}-1}$ can be obtained from $M_{4{j}-2}$ by  performing a \dop{} by~$\{{\pp^{k_j}_{j}}, {\ee^{k_j}_j}\}$.

    Analogously, $M_{4{j}-1}(c_j)={\qq^{k_j}_{j}}$ while $M_{4{j}-1}(d_j)={\ff^{k_j}_{j}}$, implying that
    $\{c_j, d_j\}$ is blocking~$M_{4{j}-1}$.
    Since ${\qq^{k_j}_{j}}$ and ${\ff^{k_j}_{j}}$ are acceptable to each other, 
    $M_{4{j}}$ can be obtained from $M_{4{j}-1}$ by  performing a \dop{} for instance by $\{c_j,d_j\}$.

    \begin{figure}[t!]
    \begin{tikzpicture}
      \node (case) {};
      \node[boy, right = 2ex of case] (pi) {};  
      \node[girl, below = \ydist of pi] (qi) {};  
      \node[boy, below = \ydist of qi] (cell) {};  
      \node[girl, below = \ydist of cell] (zell) {};  

      \node[girl, right = \xdist of pi] (ei) {};  
      \node[boy, right = \xdist of qi] (fi) {};  
      \node[girl, right = \xdist of cell] (dell) {};  
      \node[boy, right = \xdist of zell] (rell) {};  

      \foreach \i / \n / \r in {fi/{\ff^{k_j}_{j}}/right,pi/{\pp^{k_j}_{j}}/left,
        cell/{c_{j}}/left, rell/{r_{j}}/right,
        ei/{\ee^{k_j}_{j}}/right, qi/{\qq^{k_j}_{j}}/left,
        dell/{d_{j}}/right, zell/{z_{j}}/left}{
        \node[\r = 0pt of \i] (n\i) {$\n$};
      }

      \gettikzxy{(pi)}{\vx}{\vy};
      
      \gettikzxy{(ei)}{\wx}{\wy};

      \gettikzxy{(fi)}{\fx}{\fy}
      \gettikzxy{(dell)}{\dx}{\dy}
      \node at (\fx+7ex, \fy*0.5+\dy*0.5) {$\leadsto$};
      
      \node at (\vx*0.5+\wx*0.5, \vy+4ex) {$M_{4j-4}:$};
      
      \foreach \s / \t in {fi/ei, pi/qi, cell/zell, rell/dell} {
        \path%
        (\s) edge[thick] (\t);
      }
  
    \end{tikzpicture}~
    \begin{tikzpicture}
      \node (case) {};
      \node[boy, right = 2ex of case] (pi) {};  
      \node[girl, below = \ydist of pi] (qi) {};  
      \node[boy, below = \ydist of qi] (cell) {};  
      \node[girl, below = \ydist of cell] (zell) {};  

      \node[girl, right = \xdist of pi] (ei) {};  
      \node[boy, right = \xdist of qi] (fi) {};  
      \node[girl, right = \xdist of cell] (dell) {};  
      \node[boy, right = \xdist of zell] (rell) {};  

      \foreach \i / \n / \r in {fi/{\ff^{k_j}_{j}}/right,pi/{\pp^{k_j}_{j}}/left,
        cell/{c_{j}}/left, rell/{r_{j}}/right,
        ei/{\ee^{k_j}_{j}}/right, qi/{\qq^{k_j}_{j}}/left,
        dell/{d_{j}}/right, zell/{z_{j}}/left}{
        \node[\r = 0pt of \i] (n\i) {$\n$};
      }

      \gettikzxy{(pi)}{\vx}{\vy};
      
      \gettikzxy{(ei)}{\wx}{\wy};

      \gettikzxy{(fi)}{\fx}{\fy}
      \gettikzxy{(dell)}{\dx}{\dy}
      \node at (\fx+7ex, \fy*0.5+\dy*0.5) {$\leadsto$};
      
      \node at (\vx*0.5+\wx*0.5, \vy+4ex) {$M_{4j-3}:$};
      
      \foreach \s / \t / \ang in {fi/dell/0, pi/qi/0, cell/zell/0, rell/ei/-20} {
        \path%
        (\s) edge[thick, bend right=\ang] (\t);
      }
  
    \end{tikzpicture}~
    \begin{tikzpicture}
      \node (case) {};
       \node[boy, right = 2ex of case] (pi) {};  
      \node[girl, below = \ydist of pi] (qi) {};  
      \node[boy, below = \ydist of qi] (cell) {};  
      \node[girl, below = \ydist of cell] (zell) {};  

      \node[girl, right = \xdist of pi] (ei) {};  
      \node[boy, right = \xdist of qi] (fi) {};  
      \node[girl, right = \xdist of cell] (dell) {};  
      \node[boy, right = \xdist of zell] (rell) {};  

      \foreach \i / \n / \r in {fi/{\ff^{k_j}_{j}}/right,pi/{\pp^{k_j}_{j}}/left,
        cell/{c_{j}}/left, rell/{r_{j}}/right,
        ei/{\ee^{k_j}_{j}}/right, qi/{\qq^{k_j}_{j}}/left,
        dell/{d_{j}}/right, zell/{z_{j}}/left}{
        \node[\r = 0pt of \i] (n\i) {$\n$};
      }

      \gettikzxy{(pi)}{\vx}{\vy};
      
      \gettikzxy{(ei)}{\wx}{\wy};

      \gettikzxy{(fi)}{\fx}{\fy}
      \gettikzxy{(dell)}{\dx}{\dy}
      \node at (\fx+7ex, \fy*0.5+\dy*0.5) {$\leadsto$};
      
      \node at (\vx*0.5+\wx*0.5, \vy+4ex) {$M_{4j-2}:$};
      
      \foreach \s / \t / \ang in {fi/dell/0, pi/qi/0, cell/ei/0, rell/zell/0} {
        \path%
        (\s) edge[thick, bend right=\ang] (\t);
      }
    \end{tikzpicture}~
    \begin{tikzpicture}
      \node (case) {};
      \node[boy, right = 2ex of case] (pi) {};  
      \node[girl, below = \ydist of pi] (qi) {};  
      \node[boy, below = \ydist of qi] (cell) {};  
      \node[girl, below = \ydist of cell] (zell) {};  

      \node[girl, right = \xdist of pi] (ei) {};  
      \node[boy, right = \xdist of qi] (fi) {};  
      \node[girl, right = \xdist of cell] (dell) {};  
      \node[boy, right = \xdist of zell] (rell) {};  

      \foreach \i / \n / \r in {fi/{\ff^{k_j}_{j}}/right,pi/{\pp^{k_j}_{j}}/left,
        cell/{c_{j}}/left, rell/{r_{j}}/right,
        ei/{\ee^{k_j}_{j}}/right, qi/{\qq^{k_j}_{j}}/left,
        dell/{d_{j}}/right, zell/{z_{j}}/left}{
        \node[\r = 0pt of \i] (n\i) {$\n$};
      }

      \gettikzxy{(pi)}{\vx}{\vy};
      
      \gettikzxy{(ei)}{\wx}{\wy};

      \gettikzxy{(fi)}{\fx}{\fy}
      \gettikzxy{(dell)}{\dx}{\dy}
      \node at (\fx+7ex, \fy*0.5+\dy*0.5) {$\leadsto$};
      
      \node at (\vx*0.5+\wx*0.5, \vy+4ex) {$M_{4j-1}:$};
      
      \foreach \s / \t / \ang in {fi/dell/0, pi/ei/0, cell/qi/0, rell/zell/0} {
        \path%
        (\s) edge[thick, bend right=\ang] (\t);
      }
    \end{tikzpicture}~
    \begin{tikzpicture}
      \node (case) {};
      \node[boy, right = 2ex of case] (pi) {};  
      \node[girl, below = \ydist of pi] (qi) {};  
      \node[boy, below = \ydist of qi] (cell) {};  
      \node[girl, below = \ydist of cell] (zell) {};  

      \node[girl, right = \xdist of pi] (ei) {};  
      \node[boy, right = \xdist of qi] (fi) {};  
      \node[girl, right = \xdist of cell] (dell) {};  
      \node[boy, right = \xdist of zell] (rell) {};  

      \foreach \i / \n / \r in {fi/{\ff^{k_j}_{j}}/right,pi/{\pp^{k_j}_{j}}/left,
        cell/{c_{j}}/left, rell/{r_{j}}/right,
        ei/{\ee^{k_j}_{j}}/right, qi/{\qq^{k_j}_{j}}/left,
        dell/{d_{j}}/right, zell/{z_{j}}/left}{
        \node[\r = 0pt of \i] (n\i) {$\n$};
      }

      \gettikzxy{(pi)}{\vx}{\vy};
      
      \gettikzxy{(ei)}{\wx}{\wy};

      \gettikzxy{(fi)}{\fx}{\fy}
      \gettikzxy{(dell)}{\dx}{\dy}
      
      \node at (\vx*0.5+\wx*0.5, \vy+4ex) {$M_{4j}:$};
      
      \foreach \s / \t / \ang in {fi/qi/0, pi/ei/0, cell/dell/0, rell/zell/0} {
        \path%
        (\s) edge[thick, bend right=\ang] (\t);
      }
    \end{tikzpicture}
    \caption{The four \dop{s} for involving the clause-agents~$\{{\ff^{k_j}_{j}}, {\pp^{k_j}_{j}}, {\ee^{k_j}_{j}}, {\qq^{k_j}_{j}}\}$ used in the proof of \cref{thm:NP-hard-constant-length}.}\label{fig:R3SAT-dops-edge-gadget}
  \end{figure}
  After the above \dop{s}, it is straightforward to see that no blocking pairs involve any agent from~$D\cup Z\cup \{{\ff^{k_j}_{j}}, {\pp^{k_j}_{j}} \mid j \in [m]\}\cup \{{\ee^{k}_j},{\qq^{k}_j}\mid j\in [m] \wedge k\in [|e_j|]\setminus \{k_j\}\}$ as every such agent already obtains her most preferred agent.

  \item     Let $N_0=M_{4m}$.
  For each $i\in [n]$, we again perform four \dop{s} as follows:

  \noindent If $\sigma(v_i)=\true$, then define
    \begin{enumerate}[(i)]
      \item $N_{4{i}-3}=\dopm(N_{4{i}-4}, \{{\nww_{i}}, {t_{i}}\})$,
      \item $N_{4{i}-2}=\dopm(N_{4{i}-3}, \{{a_{i}}, b_{i}\})$, and
      \item $N_{4{i}-1}=\dopm(N_{4{i}-3}, \{{\nvv_{i}}, {\nyy_{i}}\})$,   
    \end{enumerate}
  \noindent  Otherwise, define
    \begin{enumerate}[(i)]
      \item $N_{4{i}-3}=\dopm(N_{4{i}-4}, \{{\ww_{i}}, {t_{i}}\})$,
      \item $N_{4{i}-2}=\dopm(N_{4{i}-3}, \{{a_{i}}, b_{i}\})$, and
      \item $N_{4{i}-1}=\dopm(N_{4{i}-2}, \{{\vvv_{i}}, {\yy_{i}}\})$.   
    \end{enumerate}
    After that, define
    \begin{enumerate}
      \item[(iv)]  $N_{4{i}}=\dopm(N_{4{i}-1}, \{t_{i}, {s_{i}}\})$.
    \end{enumerate}
    The four \dop{s} for the case that $\sigma(v_i)=\false$ are depicted in \cref{fig:dops-variable-gadget}.
      \begin{figure}[t!]
      \begin{tikzpicture}
      \node (case) {};
      \node[boy, below = 0pt of case] (yi) {};  
      \node[girl, below = \ydist of yi] (xi) {};  
      \node[boy, below = \ydist of xi] (sell) {};  
      \node[girl, below = \ydist of sell] (aell) {};

      \node[girl, right = \xdist of yi] (vi) {};  
      \node[boy, right = \xdist of xi] (wi) {};  
      \node[girl, right = \xdist of sell] (tell) {};  
      \node[boy, right = \xdist of aell] (bell) {};

      \gettikzxy{(yi)}{\firstx}{\firsty};
      \gettikzxy{(vi)}{\secondx}{\secondy};
      \gettikzxy{(wi)}{\tox}{\toy};
      \gettikzxy{(tell)}{\tox}{\tooy};
      \node at (\firstx*0.5+\secondx*0.5, \firsty+4ex) {$N_{4i-4}:$};
      \node at (\tox+7ex, \toy*0.5+\tooy*0.5) {$\leadsto$};
      
      \foreach \i / \n / \r in {vi/{\vvv_{i}}/right,xi/{\xx_{i}}/left,
        tell/{t_{{i}}}/right, aell/{a_{{i}}}/left,
        wi/{\ww_{i}}/right, yi/{\yy_{i}}/left,
        sell/{s_{{i}}}/left, bell/{b_{{i}}}/right}{
        \node[\r = 0pt of \i] (n\i) {$\n$};
      }
      
      \foreach \s / \t in {vi/wi, xi/yi, tell/bell, aell/sell} {
        \path%
        (\s) edge[thick] (\t);
      }
  
    \end{tikzpicture}~
      \begin{tikzpicture}
      \node (case) {};
      \node[boy, below = 0pt of case] (yi) {};  
      \node[girl, below = \ydist of yi] (xi) {};  
      \node[boy, below = \ydist of xi] (sell) {};  
      \node[girl, below = \ydist of sell] (aell) {};

      \node[girl, right = \xdist of yi] (vi) {};  
      \node[boy, right = \xdist of xi] (wi) {};  
      \node[girl, right = \xdist of sell] (tell) {};  
      \node[boy, right = \xdist of aell] (bell) {};

      \gettikzxy{(yi)}{\firstx}{\firsty};
      \gettikzxy{(vi)}{\secondx}{\secondy};
      \gettikzxy{(wi)}{\tox}{\toy};
      \gettikzxy{(tell)}{\tox}{\tooy};
      \node at (\firstx*0.5+\secondx*0.5, \firsty+4ex) {$N_{4i-3}:$};
      \node at (\tox+7ex, \toy*0.5+\tooy*0.5) {$\leadsto$};
      
      \foreach \i / \n / \r in {vi/{\vvv_{i}}/right,xi/{\xx_{i}}/left,
        tell/{t_{{i}}}/right, aell/{a_{{i}}}/left,
        wi/{\ww_{i}}/right, yi/{\yy_{i}}/left,
        sell/{s_{{i}}}/left, bell/{b_{{i}}}/right}{
        \node[\r = 0pt of \i] (n\i) {$\n$};
      }
      
      \foreach \s / \t / \ang in {vi/bell/-20, tell/wi/0, sell/aell/0, xi/yi/0} {
        \path%
        (\s) edge[thick, bend left = \ang] (\t);
      }
    \end{tikzpicture}~
    \begin{tikzpicture}
      \node (case) {};
      \node[boy, below = 0pt of case] (yi) {};  
      \node[girl, below = \ydist of yi] (xi) {};  
      \node[boy, below = \ydist of xi] (sell) {};  
      \node[girl, below = \ydist of sell] (aell) {};

      \node[girl, right = \xdist of yi] (vi) {};  
      \node[boy, right = \xdist of xi] (wi) {};  
      \node[girl, right = \xdist of sell] (tell) {};  
      \node[boy, right = \xdist of aell] (bell) {};

      \gettikzxy{(yi)}{\firstx}{\firsty};
      \gettikzxy{(vi)}{\secondx}{\secondy};
      \gettikzxy{(wi)}{\tox}{\toy};
      \gettikzxy{(tell)}{\tox}{\tooy};
      \node at (\firstx*0.5+\secondx*0.5, \firsty+4ex) {$N_{4i-2}:$};
      \node at (\tox+7ex, \toy*0.5+\tooy*0.5) {$\leadsto$};
      
      \foreach \i / \n / \r in {vi/{\vvv_{i}}/right,xi/{\xx_{i}}/left,
        tell/{t_{{i}}}/right, aell/{a_{{i}}}/left,
        wi/{\ww_{i}}/right, yi/{\yy_{i}}/left,
        sell/{s_{{i}}}/left, bell/{b_{{i}}}/right}{
        \node[\r = 0pt of \i] (n\i) {$\n$};
      }
      
      \foreach \s / \t / \ang in {vi/sell/0, aell/bell/0, tell/wi/0, xi/yi/0} {
        \path%
        (\s) edge[thick, bend left = \ang] (\t);
      } 
    \end{tikzpicture}~
    \begin{tikzpicture}
      \node (case) {};
      \node[boy, below = 0pt of case] (yi) {};  
      \node[girl, below = \ydist of yi] (xi) {};  
      \node[boy, below = \ydist of xi] (sell) {};  
      \node[girl, below = \ydist of sell] (aell) {};

      \node[girl, right = \xdist of yi] (vi) {};  
      \node[boy, right = \xdist of xi] (wi) {};  
      \node[girl, right = \xdist of sell] (tell) {};  
      \node[boy, right = \xdist of aell] (bell) {};

      \gettikzxy{(yi)}{\firstx}{\firsty};
      \gettikzxy{(vi)}{\secondx}{\secondy};
      \gettikzxy{(wi)}{\tox}{\toy};
      \gettikzxy{(tell)}{\tox}{\tooy};
      \node at (\firstx*0.5+\secondx*0.5, \firsty+4ex) {$N_{4i-1}:$};
      \node at (\tox+7ex, \toy*0.5+\tooy*0.5) {$\leadsto$};
      
      \foreach \i / \n / \r in {vi/{\vvv_{i}}/right,xi/{\xx_{i}}/left,
        tell/{t_{{i}}}/right, aell/{a_{{i}}}/left,
        wi/{\ww_{i}}/right, yi/{\yy_{i}}/left,
        sell/{s_{{i}}}/left, bell/{b_{{i}}}/right}{
        \node[\r = 0pt of \i] (n\i) {$\n$};
      }
      
      \foreach \s / \t / \ang in {vi/yi/0, aell/bell/0, tell/wi/0, xi/sell/0} {
        \path%
        (\s) edge[thick, bend left = \ang] (\t);
      } 
    \end{tikzpicture}~
   \begin{tikzpicture}
      \node (case) {};
      \node[boy, below = 0pt of case] (yi) {};  
      \node[girl, below = \ydist of yi] (xi) {};  
      \node[boy, below = \ydist of xi] (sell) {};  
      \node[girl, below = \ydist of sell] (aell) {};

      \node[girl, right = \xdist of yi] (vi) {};  
      \node[boy, right = \xdist of xi] (wi) {};  
      \node[girl, right = \xdist of sell] (tell) {};  
      \node[boy, right = \xdist of aell] (bell) {};

      \gettikzxy{(yi)}{\firstx}{\firsty};
      \gettikzxy{(vi)}{\secondx}{\secondy};
      \gettikzxy{(wi)}{\tox}{\toy};
      \gettikzxy{(tell)}{\tox}{\tooy};
      \node at (\firstx*0.5+\secondx*0.5, \firsty+4ex) {$N_{4i}:$};
      
      \foreach \i / \n / \r in {vi/{\vvv_{i}}/right,xi/{\xx_{i}}/left,
        tell/{t_{{i}}}/right, aell/{a_{{i}}}/left,
        wi/{\ww_{i}}/right, yi/{\yy_{i}}/left,
        sell/{s_{{i}}}/left, bell/{b_{{i}}}/right}{
        \node[\r = 0pt of \i] (n\i) {$\n$};
      }
      
      \foreach \s / \t / \ang in {vi/yi/0, aell/bell/0, xi/wi/0, tell/sell/0} {
        \path%
        (\s) edge[thick, bend left = \ang] (\t);
      } 
    \end{tikzpicture} 
    \caption{The four \dop{s} for involving the vertex-agents~$\{{\vvv_{i}}, {\xx_{i}}, {\ww_{i}}, {\yy_{i}}\}$ when $\sigma(v_i)=\false$ used in the proof of \cref{thm:NP-hard-constant-length}.}\label{fig:dops-variable-gadget} 
    \end{figure} 
    After the above \dop{s}, it is straightforward to see that no blocking pairs involve any agent from~$T\cup A\cup \{{\nvv_i}, {\nxx_{i}}, {\ww_i}, {\yy_i} \mid v_i\in V \wedge \sigma(v_i)=\true\}\cup
    \{{\vvv_i}, {\xx_{i}}, {\nww_i}, {\nyy_i}\mid v_i\in V \wedge \sigma(v_i)=\false\}$ as every such agent already obtains her most preferred agent. 
  \end{enumerate}
  In total, we have performed~$4m+4n$~\dop{s} resulting in a valid matching~$N_{4n}$.
  The proof that $N_{4n}$ is stable is analogous to the one given in \cref{thm:divor-incomplete-noties-nphard}.
  To ease notation, let $M\coloneqq N_{4n}$.
  Since we are in the bipartite case, to show stability, we will show that no blocking pair involves an agent from $\hat{W}$.
   More precisely, we show the following:
  \begin{itemize}[--]
    \item Consider an arbitrary~$i\in [n]$. By our sequence of \dop{s},
    if $\sigma(v_i)=\true$, then $M({\nww_i}) = {\nxx_i}$ and $M({\nyy_i})={\nvv_i}$.
    Hence, no agent from $\{{\nww_i}, {\nyy_i}\}$ is involved in a blocking pair.
    Furthermore, $M({\ww_i}) = {\vvv_i}$, $M({\yy_i})={\xx_i}$, $M(t_i)=s_i$, $M(a_i)=b_i$
    This means that no agent~$\phi$ from $\{{\ww_i}, {\yy_i}, b_i\}$ is involved in a blocking pair since for each agent~$\psi$ that $\phi$ prefers to her partner~$M(\phi)$ it holds that $M(\psi)$ is the most preferred partner of~$\psi$.
    Similarly, we can verify that neither can $s_i$ be involved in a blocking pair.
    The case when $\sigma(v_i)=\false$ can be shown analogously.
    \item Consider an arbitrary~$j \in [m]$. Recall that $k_j$ was defined as an index such that the $k_j^{\text{th}}$ literal of~$e_j$ is set $\true$ under~$\sigma$.
    The only clause-agents that we need to consider are ${\ee^{k_j}_j}$ and ${\qq^{k_j}_j}$ since all other clause-agents for~$e_j$ obtain their most preferred agents.
    Clearly, ${\qq^{k_j}_j}$ is not involved in any blocking pair since she only prefers~$\pp^{k_j}_{j}$ and $c_j$ to her partner~${\pp^{k_j}_j}$ but none of $\{{\pp^{k_j}_j}, c_j\}$ prefers~$\qq^{k_j}_j$ to her respective partner.
    As for~$\ee^{k_j}_j$, whose partner is~$\pp^{k_j}_j$, observe that there are three agents, namely $\ff^{k_j}, r_j$, and $v({\ee^{k_j}_j})$, which $\ee^{k_j}_j$ prefers to~$\pp^{k_j}_j$.
    But none of those prefers to be with $\ee^{k_j}_j$ due to the following.
    \begin{itemize}[--]
      \item[$\bullet$] $\ff^{k_j}_j$ already obtains her most preferred partner.
      \item[$\bullet$] $r_j$ has partner~$z_j$ and she prefers $z_j$ to~$\ee^{k_j}_j$.
      \item[$\bullet$] $v({\ee^{k_j}_j})$ corresponds to the literal which satisfies~$e_j$ and by our definition of the \dop{s}, agent~$v({\ee^{k_j}_j})$ remain obtaining her most preferred agent.
    \end{itemize}
  \end{itemize}
  Since no agent from~$\hat{W}$ is involved in a blocking pair, the reachable matching~$M$ is indeed stable.

    For the ``if'' part of the correctness proof, assume that there exists a stable matching, denoted as~$M_\ell$, which is reachable from~$M_0$.
    Let $L'=(\rho_0,\rho_1,\ldots,\rho_{\ell-1})$ be such a witness for~$M_\ell$ to be reachable from~$M_0$.
    Before we show how to construct a satisfying truth assignment, 
    we observe that for each~$j\in [m]$ each clause-selector-agent~$d_j$ (resp.\ $z_j$) will help exactly one of the clause-agents~$\{{\ff^{k}_j}\mid k\in [|e_j|]\}$ reaching her most preferred agent, namely~${\qq_j}$.
    Let this agent be~$\ff^{k_j}_j$.
    Then, by \cref{claim:R3SAT-SM-prop}\eqref{R3SAT-SM-prop:E}, agent~${\ee^{k_j}_j}$ will need to be matched to~${\pp^{k_j}_j}$.
    By the preferences of $\ee^{k_j}_j$ and its corresponding literal-agent~$v({\ee^{k_j}_j})$
    and by \cref{claim:R3SAT-SM-prop}\eqref{R3SAT-SM-prop:SV},
    it follows that $M(v({\ee^{k_j}}_j)) \in W$.
    Setting the literal corresponding to $v({\ee^{k_j}_j})$ to $\true$ gives us a satisfying assignment.

  We formalize the above idea through the following technical properties for the sequence~$L'$.
  \begin{claim}\label{claim:R3SAT-reachable-SM}
    For each~$i\in [\ell]$, define $M_{i}\coloneqq \dopm(M_{i-1}, \rho_{i-1})$.
    Then, the following holds.
    \begin{enumerate}[(1)]

      \item\label{R3SAT-reachable-SM-T1}
      For each~$i\in [n]$,
      there exist a pair~$({\vvv}, {\ww}) \in \{({\vvv_i}, {\ww_i}), ({\nvv_i}, {\nww_i})\}$ and an index~$\alpha\in [\ell]$ such that
      $M_{\alpha-1}(t_i) = b_i$, $M_{\alpha-1}({\vvv}) = {\ww}$, 
      $M_{\alpha}(t_i) = {\ww}$, and $M_{\alpha}({\vvv}) = {b_i}$.

        \item \label{R3SAT-reachable-SM-D0-unique} For each $d_j\in D$ it holds that $M_{\ell}(d_{j})=c_{j}$.
      \item\label{R3SAT-reachable-SM-D1}
      For each~$j\in [m]$,
      there exist two indices~$k_j\in [|e_j|]$ and~$\alpha\in [\ell]$ such that
      $M_{\alpha-1}(d_j) = r_j$, $M_{\alpha-1}({\ee^{k_j}_j}) = {\ff^{k_j}_j}$,
      $M_{\alpha}(d_j) = {\ff^{k_j}_j}$, and $M_{\alpha}({\ee^{k_j}_j}) = {r_j}$.

    \end{enumerate}
  \end{claim}

  \iflong
  \begin{proof} \renewcommand{\qedsymbol}{(of \cref{claim:R3SAT-reachable-SM})~\cqed}
    \noindent\textbf{Statement~\eqref{R3SAT-reachable-SM-T1}:}
    The proof for this statement follows the same line as the one for \cref{claim:reachable-SM}\eqref{reachable-SM-D1}; we repeat for the sake of completeness.
    Consider an arbitrary~$i\in [n]$.
    We first show that $M_{\ell}(t_i)=s_i$.
    Suppose, for the sake of contradiction, that $M_{\ell}(t_i)\neq s_i$.
    By \cref{claim:R3SAT-SM-prop}\eqref{R3SAT-SM-prop:SV}, it follows that $M_{\ell}(t_i) = b_i$.
    Since $t_i$ prefers both ${\ww_i}$ and ${\nww_i}$ to $b_i$,
    by the stability of~$M$, it follows that $M_{\ell}({\ww_i})={\xx_i}$ and $M_{\ell}({\nww_i})={\nxx_i}$.

    For agent~$\ww_i$, by the initial matching~$M_0$, there must be two consecutive matchings where the partner of $\ww_i$ changes from someone other than~$\xx_i$ to~$\xx_i$.
    Let $M_{\alpha-1}$ and $M_{\alpha}$, $\alpha\in [\ell]$, be two consecutive matchings witnessing this, i.e., $M_{\alpha-1}({\ww_i})\neq {\xx_i}$ and $M_{\alpha}({\ww_i})={\xx_i}$.
    Since $M_{\alpha}=\dopm(M_{\alpha-1}, \rho_{\alpha-1})$, by the definition of \dop{s},
    it follows that
    $\rho_{\alpha-1}=\{{\ww_i}, {\xx_i}\}$ or $\rho_{\alpha-1}=\{M_{\alpha-1}({\ww_i}), M_{\alpha-1}({\xx_i})\}$.
    Since ${\ww_i}$ is the least preferred agent of ${\xx_i}$,
    we infer that $\rho_{\alpha-1}=\{M_{\alpha-1}({\ww_i}), M_{\alpha-1}({\xx_i})\}$.
    Since ${\ww_i}$ is the most preferred agent of~${\vvv_i}$,
    we also infer that $M_{\alpha-1}({\ww_i})\neq {\vvv_i}$.
    By the acceptable partners of~${\ww_i}$ it follows that $M_{\alpha-1}({\ww_i})=t_i$.
    Observe that besides~${\ww_i}$ only~$s_i$ finds both~${\ww_i}$ and $\ww_i$ acceptable.
    This implies that $M_{\alpha-1}({\xx_i})=s_i$ and $M_{\alpha}(t_i)=s_i$.
    By \cref{claim:R3SAT-reachable-matching}\eqref{R3SAT-reachable-sk}, $t_{i}$ and $s_{i}$ remain matched to each other in $(M_{\alpha}, M_{\alpha+1}, \ldots, M_{\ell})$, a contradiction to our assumption that $M_{\ell}(t_i)\neq s_i$.

    We have just shown that $M_{\ell}(t_i)=s_i$.
    Observe that the above proof already reveals how to find such a pair~$(v,w)$ for the statement.
    First of all, since $M_{\ell}(t_i)=s_i$, by the initial matching, the partner of~$t_i$ changes from $b_i$ to someone else.
    Let $M_{\alpha-1}$ and $M_{\alpha}$ in~$L$, $\alpha\in [\ell]$, be two consecutive matchings
    such that $M_{\alpha-1}(t_i) = b_i$ while $M_{\alpha}(t_i)\neq b_i$.
    Since $M_{\alpha}=\dopm(M_{\alpha-1}, \rho_{\alpha-1})$,
    it follows that
    $\rho_{\alpha-1}=\{t_i, M_{\alpha}(t_i)\}$ or $\rho_{\alpha-1}=\{b_i, M_{\alpha}(b_i)\}$.
    Since $t_i$ is $b_i$'s most preferred agent, 
    we infer that $\rho_{\alpha-1}=\{t_i, M_{\alpha}(t_i)\}$.
    By the acceptable agents of $b_i$, it follows that $M_{\alpha}(b_i)\in \{a_j,{\vvv_i},{\nvv_i}\}$.
    Since $M_{\alpha}(b_i)$ and $M_{\alpha}(t_i)$ are matched under~$M_{\alpha-1}$, we infer that $M_{\alpha}(b_i)\neq a_i$ since no agent, except~$b_i$, prefers $t_i$ to~$a_j$ so
    that $\{t_i, M_{\alpha}(t_i)\}$ cannot be blocking~$M_{\alpha-1}$.
    This means that $M_{\alpha}(b_i) \in \{{\vvv_i}, {\nvv_i}\}$.
    Since, except~$b_i$, agent~${\ww_i}$ (resp.\ ${\nww_i}$) is the only agent who prefers~$t_i$ to $\vvv_i$ (resp.\ ${\nvv_i}$),
    we infer that either $M_{\alpha}(t_i)={\ww_i}$ and $M_{\alpha}(b_i)={\vvv_i}$ or
    $M_{\alpha}(t_i)={\nww_i}$ and $M_{\alpha}(b_i)={\nvv_i}$.
    Summarizing, we have found a pair~$(v,w)\in \{({\vvv_i}, {\ww_i}), ({\nvv_i}, {\nww_i})\}$ and an index~$\alpha\in [\ell]$ for the statement.
    
     \noindent\textbf{Statement~\eqref{R3SAT-reachable-SM-D0-unique}:}
    The proof for this statement follows the same line as the one for \cref{claim:reachable-SM}\eqref{reachable-SM-D0-unique}; we repeat for the sake of completeness.
    Suppose, for the sake of contradiction, that there exists a clause-selector-agent~$d_j\in D$ with $M_{\ell}(d_j)\neq c_j$.
    By \cref{claim:R3SAT-SM-prop}\eqref{R3SAT-SM-prop:SE}, it follows that $M_{\ell}(d_j) = r_j$.
    Then, by the preferences of the $F$-agents and by \cref{claim:R3SAT-SM-prop}\eqref{R3SAT-SM-prop:E},
    for each $k\in [|e_j|]$, it must hold that $M_{\ell}({\ff^{k}_j}) = {\qq^{k}_j}$ as otherwise~$\{{\ff^k_j}, d_j\}$ would be blocking~$M_\ell$.
    Consider an arbitrary clause-agent~${\ff^k_j}$, $k\in [|e_j|]$.
    Since $M_0({\ff^k_j})={\ee^k_j}\neq {\qq^k_j}$,
    there exist two consecutive matchings in~$L$ where the partner of~${\ff^k_j}$ changes from someone other than~${\qq^k_j}$ to~${\qq^k_j}$.
    Let $M_{\alpha-1}$ and $M_{\alpha}$, $\alpha\in [\ell]$, be two consecutive matchings witnessing this, i.e., $M_{\alpha-1}({\ff^k_j})\neq {\qq^k_j}$ and $M_{\alpha}({\ff^k_j})={\qq^k_j}$.
    Since $M_{\alpha}=\dopm(M_{\alpha-1}, \rho_{\alpha-1})$, by the definition of \dop{s},
    it follows that
    $\rho_{\alpha-1}=\{{\ff^k_j}, {\qq^k_j}\}$ or $\rho_{\alpha-1}=\{M_{\alpha-1}({\ff^k_j}), M_{\alpha-1}({\ff^k_j})\}$.
    Since ${\ff^k_j}$ is the least preferred agent of ${\qq^k_j}$,
    we infer that $\rho_{\alpha-1}=\{M_{\alpha-1}({\ff^k_j}), M_{\alpha-1}({\qq^k_j})\}$.
    Since ${\ff^k_j}$ is the most preferred agent of~${\ee^k_i}$,
    we also infer that $M_{\alpha-1}({\ff^k_j})\neq {\ee^k_j}$.
    By the acceptable partners of~${\ff^k_j}$ it follows that $M_{\alpha-1}({\ff_j})=d_{j}$.
    Observe that besides~${\ff^k_j}$ only~$c_j$ finds both~${\qq^k_j}$ and $d^k_j$ acceptable.
    This implies that $M_{\alpha-1}({\qq^k_j})=c_j$ and $M_{\alpha}(d_j)=c_j$.
    By \cref{claim:R3SAT-reachable-matching}\eqref{R3SAT-reachable-ck}, $d_{k}$ and $c_{k}$ remain matched to each other in $(M_{\alpha}, M_{\alpha+1}, \ldots, M_{\ell})$, a contradiction to our assumption that $M_{\ell}(d_j)\neq r_j$.
    
    \noindent\textbf{Statement~\eqref{R3SAT-reachable-SM-D1}:} 
    The proof for this statement follows the same line as the one for \cref{claim:reachable-SM}\eqref{reachable-SM-D1}; we repeat for the sake of completeness.
   Consider an arbitrary $j\in [m]$.
    By Statement~\eqref{R3SAT-reachable-SM-D0-unique} and since $M_0(d_j)=r_j$,
    there must be two consecutive matchings~$M_{\alpha-1}$ and $M_{\alpha}$ in~$L$, $\alpha\in [\ell]$,
    such that $M_{\alpha-1}(d_j) = r_j$ while $M_{\alpha}(d_j)\neq r_j$.
    Since $M_{\alpha}=\dopm(M_{\alpha-1}, \rho_{\alpha-1})$,
    it follows that
    $\rho_{\alpha-1}=\{d_j, M_{\alpha}(d_j)\}$ or $\rho_{\alpha-1}=\{r_j, M_{\alpha}(r_j)\}$.
    Since $d_j$ is $r_j$'s most preferred agent, 
    we infer that $\rho_{\alpha-1}=\{d_j, M_{\alpha}(d_j)\}$.
    By the acceptable agents of $r_j$, it follows that $M_{\alpha}(r_j)\in \{z_j\}\cup \{{\ee^k_j}\mid k\in [|e_j|]\}$.
    Since $M_{\alpha}(d_j)$ and $M_{\alpha}(r_j)$ are matched under~$M_{\alpha-1}$, we infer that $M_{\alpha}(r_j)\neq z_j$ since no agent, except~$d_j$, prefers $d_j$ to~$z_j$ so
    that $\{d_j, M_{\alpha}(d_j)\}$ cannot be blocking~$M_{\alpha-1}$.
    This means that $M_{\alpha}(r_j) = {\ee^{k_j}_j}$ for some $k_j\in [|e_j|]$.
    Since, except~$r_j$, agent~${\ff^{k_j}_j}$ is the only agent who prefers~$d_j$ to $\ee^{k_j}_j$,
    we infer that $M_{\alpha}(d_j)={\ff^{k_j}_j}$.
    Summarizing, we have found such two indices~$k_j\in [|e_j|]$ and $\alpha\in [\ell]$ for the statement.
  \end{proof}
  \fi

  Now, we show that $I$ admits a satisfying truth assignment.
  For each~$j\in [m]$, let
  $k_j\in [|e_j|]$ and $\alpha_j\in [\ell]$ denote the two indices according to \cref{claim:R3SAT-reachable-SM}\eqref{R3SAT-reachable-SM-D1}.
  We claim that the following assignment~$\sigma\colon V\to \{\true,\false\}$ is a satisfying assignment:
  \begin{itemize}[--]
    \item For each $j\in [m]$ if $\ee^{k_j}_j$ corresponds to some un-negated literal~$v_i\in V$,
  then let $\sigma(v_i)=\true$; otherwise, meaning that $\ee^{k_j}_j$ corresponds to some negated literal~$\overline{v}_i \in V$, then let $\sigma(v_i)=\false$.
  \item Assign the remaining not-yet-considered variables arbitrarily, for instance, to~$\true$.
\end{itemize}
We first show that $\sigma$ is a valid assignment, i.e., there exist no two clause agents~$\ee^{k_j}_j$ and $\ee^{k_{j'}}_{j'}$ which correspond to the un-negated and negated literals of the same variable.
Suppose, for the sake of contradiction, that $\ee^{k_j}_j$ and $\ee^{k_{j'}}_{j'}$ correspond $v_i$ and $\overline{v}_i$, respectively, for some~$i\in [n]$.
By \cref{claim:R3SAT-reachable-SM}\eqref{R3SAT-reachable-SM-D1} (applying to~$j$ and $j'$),
this means that $M_{\alpha_j}({\ee^{k_j}_j})=r_j$ and  $M_{\alpha_j}({\ee^{k_{j'}}_{j'}})=r_{j'}$.
By \cref{claim:R3SAT-reachable-matching}\eqref{R3SAT-reachable-e} and \cref{claim:R3SAT-SM-prop}\eqref{R3SAT-SM-prop:E}  we infer that $M({\ee^{k_j}_j}) = {\pp^{k_j}_j}$ and $M({\ee^{k_{j'}}_{j'}}) = {\pp^{k_{j'}}_{j'}}$.
Since $M({\ee^{k_j}_j})$  and $\ee^{k_{j'}}_{j'}$ prefer $\vvv_i$ and $\nvv_i$ to their own partners, respectively,  
by the preferences of~$\vvv_i$ and $\nvv_i$ and by \cref{claim:R3SAT-SM-prop}\eqref{R3SAT-SM-prop:V}--\eqref{R3SAT-SM-prop:negV}, this implies that $M({\vvv_i})={\ww_i}$ and $M({\nvv_i})={\nww_i}$.
By the contra-positive of \cref{claim:R3SAT-reachable-matching}\eqref{R3SAT-reachable-v},
this means that during the whole sequence, the partner of ${\ww_i}$ and ${\nww_i}$ remain unchanged, a contradiction to \cref{claim:R3SAT-reachable-SM}\eqref{R3SAT-reachable-SM-T1}.

Now, we show that $\sigma$ satisfies every clause.
Let $e_j\in E$ be an arbitrary clause.
By our definition of $k_j$ and the reasoning above, it follows that $M({\ee^{k_j}_j})={\pp^{k_j}_j}$.
Since we define the truth value of the $k_j^{\text{th}}$ literal in $e_j$ according to the literal-agent~$\ee^{k_j}_j$, it follows that $e_j$ is satisfied by the $k_j^{\text{th}}$ literal. 
\fi
\end{proof}

\subsection{Combining $\boldsymbol{\kappa}$ with~$\boldsymbol{d}$}
For the combined parameter~``max.\ number~$\kappa$ of allowed divorces'' and ``max.\ preferences length~$d$'',
we obtain fixed-parameter tractability, using the following observation.

\begin{lemma}\label{lem:max-bps}
  For each yes-instance~$I=(U,W,(\succ_x)_{x\in U\cup W}, M_0)$ of \divorcesms{}, %
  the number of blocking pairs of~$M_0$ is at most~$4(d-1)\kappa$,
  $d$ denotes the maximum length of the preferences and $\kappa$ the length of the shortest witness for~$M_0$.
\end{lemma}

\begin{proof}
  To show the statement, we only need to observe that each divorce operation changes the partners of four agents,
  and hence can reduce the number of blocking pairs by at most~$4(d-1)$.
  Since after~$\kappa$ divorce operations we arrive at a stable matching, meaning that the number of blocking pairs drops to zero,
  the number of blocking pairs of the initial matching~$M_0$ is at most~$4(d-1)\kappa$.
\end{proof}

\noindent Using \cref{lem:max-bps}, we can prove \cref{thm:FPT}.

\iflong \begin{proof}[Proof of \cref{thm:FPT}] \else  \begin{proof}[Proof sketch of \cref{thm:FPT}] \fi
  The idea is to adapt the XP algorithm for \cref{cor:dop-xp} and use \cref{lem:max-bps} to additionally check whether the number of blocking pairs is bounded.
  \ifshort The approach is described in \cref{alg:fpt-k+d} (see the \callMain{} function in Line~\ref{line:main}).
  \else

  Let $I=(U,W,(\succ_x)_{x\in U\cup W}, M_0)$ be an instance of \divorcesms{}, with $|U|=|W|=n$,
  and $d$ denote the maximum length of the preferences and $\kappa$ the length of the shortest witness for~$M_0$.
  \iflong  We need one more notion: \fi
  For each matching~$N$ of $I$, let \myemph{$\bps(N)$} denote the set of blocking pairs of $N$.
  
  Using \cref{lem:max-bps}, we branch into all $4(d-1)\kappa$ possible divorces defined by the blocking pairs of a current matching and check whether after $\kappa$ iterations of branches at least one branch leads to a witness for~$M_0$.
  The procedure is described in \cref{alg:fpt-k+d} (see the \callMain{} function in Line~\ref{line:main}).
  The correctness follows directly from \cref{lem:max-bps}. %

  \fi  
  As for the running time, computing the set of blocking pairs of a matching can be done in $O(n^2)$ time.
  Starting with $i=1$, for each call to \callcheckBP{} with argument~($M$, $i$, $\kappa$, $d$) ($M$ denotes the current matching to be considered and $i$ the length of the sequence of the matchings considered so far),
  the number of recursive calls is at most~$4(d-1)(\kappa+1-i)$~(see Line~\ref{line:recurse}),
  We stop when the length of the sequence of the considered matchings for each branch reaches~$\kappa$; note that we use~$i$ to store the length of the sequence.
  Hence, the total running time is $O(n^2\cdot \prod_{i=1}^{\kappa}(4d-4)\cdot (\kappa+1-i)) = O(n^2\cdot (4d)^{\kappa}\cdot \kappa!)$, as desired.
\end{proof}

\begin{algorithm}[t!]
  \DontPrintSemicolon
  \caption{A branching algorithm for \cref{thm:FPT}}\label[algorithm]{alg:fpt-k+d}
  \small
  \SetKwFunction{checkBP}{checkBP}
  \SetKwFunction{main}{main}

  \SetKwProg{Fn}{Function}{:}{}

  \label{line:main}  \lFn{\main{$U,W,(\succ_x)_{x\in U\cup W}, M_0, \kappa, d$}}{
    \Return \checkBP($M_0$, $1$, $\kappa$, $d$)}
  
  \Fn{\checkBP{$M$, $i$, $\kappa$, $d$}}
  {
    \lIf{$i> \kappa$ {\normalfont\textbf{or}} $|\bps(M)| > 4(d-1)(\kappa+1-i)$}{%
      \Return $\false$
    }
    
    \lIf{$\bps(M)=\emptyset$}{\Return $\true$}
    
     \lForEach{$\rho\in \bps(M)$}{
      \lIf*{\normalfont \checkBP($i+1$, $\dopm(M,\rho)$, $\kappa$, $d$)}{\Return $\true$}    \label{line:recurse}}
    
    \Return $\false$
  }

\end{algorithm}

\section{Open questions}\label{sec:conclude}
Our work leads to several open questions.
First of all, the most pressing question is whether the problem remains intractable when the preferences are complete.
Secondly, to answer the second question of Knuth concerning the length of the shortest witness, it is important to know whether the problem is actually contained in NP.

\subsubsection*{Acknowledgments}
I thank Du\v{s}an Knop (Czech Technical University in Prague, Czech Republic) and Junjie Luo (TU Berlin, Germany) for initial discussion on the project. 
I am also grateful to Girija Limaye (IIT Madras, India) who helped to improve on the 2nd version of the paper.
I was supported by the European Research Council (ERC) under the European Union’s Horizon 2020 research and innovation programme under grant agreement numbers~677651.\\ \includegraphics[width=50px]{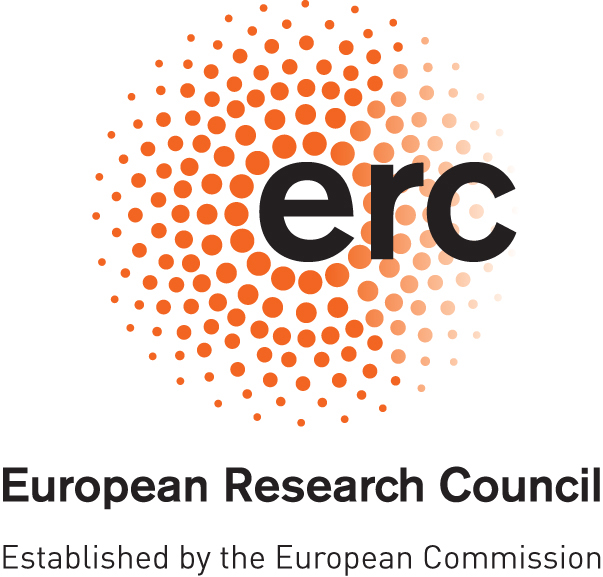}\hspace{.5cm} \includegraphics[width=50px]{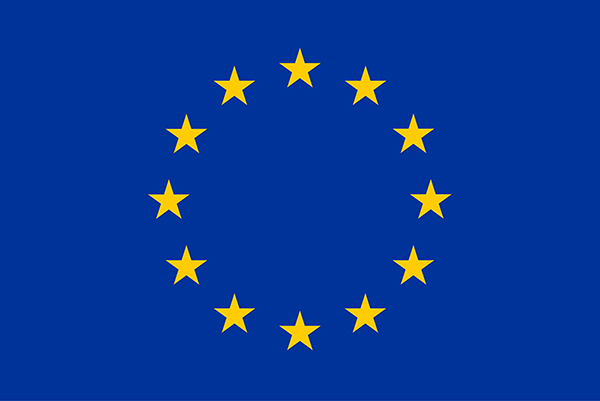}%
\\
I am also supported by the WWTF research project ``Structural and Algorithmic Aspects of Preference-based Problems in Social Choice'' (VRG18-012).

\bibliographystyle{abbrvnat}

\bibliography{bib}

\begin{thebibliography}{17}
\bibitem[Cheng(2016)]{Cheng2016}
C.~T. Cheng.
\newblock On the stable matchings that can be reached when the agents go
  marching in one by one.
\newblock \emph{SIAM Journal on Discrete Mathematics}, 30\penalty0
  (4):\penalty0 2047--2063, 2016.
\end{thebibliography}

\end{document}